\documentclass{astro-ph-elsarticle}

\newcommand {\0}  {\phantom{0}}
\newcommand {\da} {\ensuremath{\Delta a_0}}
\newcommand {\dr} {\ensuremath{\Delta r}}
\newcommand {\fm} {\ensuremath{\mathit{f}_\mathcal{M}}}

\usepackage{lineno,hyperref}
\modulolinenumbers[5]

\journal{Planetary and Space Science}

\biboptions{authoryear}

\begin{document}

\newpage
\setcounter{page}{1}

\begin{frontmatter}

\title{Evidence of Eta Aquariid Outbursts Recorded in the Classic Maya Hieroglyphic Script Using Orbital Integrations}

\author{J. H. Kinsman\fnref{myfootnote}}
\address{6324 Chesla Dr, Gainesville, Georgia, 30506, USA}
\fntext[myfootnote]{email address (corresponding author):  jhkinsman@gatech.edu}

\author{D. J. Asher\fnref{DJAemail}}
\address{Armagh Observatory \& Planetarium, College Hill, Armagh, BT61 9DG, UK}
\fntext[DJAemail]{email address:  dja@arm.ac.uk}

\begin{abstract}
No firm evidence has existed that the ancient Maya civilization recorded
specific occurrences of meteor showers or outbursts in the corpus of Maya
hieroglyphic inscriptions.  In fact, there has been no evidence of any
pre-Hispanic civilization in the Western Hemisphere recording any
observations of any meteor showers on any specific dates.

The authors numerically integrated meteoroid-sized particles released by
Comet Halley as early as 1404 BC to identify years within the Maya Classic
Period, AD 250--909, when Eta Aquariid outbursts might have occurred.
Outbursts determined by
computer model were then compared to specific events in the Maya record to
see if any correlation existed between the date of the event and the date of
the outburst.  The model was validated by successfully explaining
several outbursts around the same epoch in the Chinese record.
Some outbursts observed by the Maya were due to recent
revolutions of Comet Halley, within a few centuries, and some to resonant
behavior in older Halley trails, of the order of a thousand years.  Examples
were found of
several different Jovian mean motion resonances as well as the 1:3 Saturnian
resonance that have controlled the dynamical evolution of meteoroids in
apparently observed outbursts.
\end{abstract}

\begin{keyword}
\texttt{Maya astronomy, archaeoastronomy, meteor outburst, Eta Aquariids}
\end{keyword}

\end{frontmatter}

\large 


\section{Introduction}

\subsection{Historical background}

Investigating meteor outbursts or even meteor showers at all in the Maya
records presents unique problems since a majority of the ancient books known
as codices, possibly containing original astronomical observations, were destroyed by the
Spanish after their arrival into Maya territory in the 16th 
century\footnote{In 1562 under the direction of Bishop Diego de Landa, a large
number of codices were burned in an action known as the \textit{auto de fe}
 \cite[p.~169, 77, 134]{landa66}.}.  Surviving tables and almanacs  
 in these books contain astronomical information
 relating to Venus, solar and lunar eclipses, and seasonal information for agricultural
 purposes.  Stone monuments, panels, painted murals and portable 
objects such as bones, shells and ceramic vases however
still do exist from the Classic Period and contain hieroglyphic inscriptions 
that record close to an estimated 2000\footnote{Counts separately 
dates duplicated at different sites.} dates in the Maya calendar
\citep{mathews16}.  
Many of the dates carved in stone record dynastic information such 
as lineage,  births,  accessions to rulership and deaths, war events such as ``axing,'' 
``prisoner-capture'' and ``Star War'' victories over rival polities, and dedicatory events
such as Period Endings (see footnote \ref{note-pd-ending}) and fire ceremonies.
Although much of the inital information inscribed on stelae includes lunar information
 such as the age of the moon, and the number and length of the lunation 
\citep[see for instance][]{sgf92}, little else seemed to have been inscribed outright regarding astronomical 
 information\footnote{\label{note-eclipse} One notable exception of a recorded astronomical event 
 during the Classic Period is a solar eclipse of AD 790 found on a monument at
  the site of Santa Elena Poco Uinic.}. Incredibly, that notion changed in 2012 with
the discovery of an early 9th century astronomer's workshop \citep{ssar12} that contained lunar
  tables and numbered arrays painted on the walls of a small room indicating commensuration 
  applications to various Maya calendrical and astronomical cycles.  
  
  Clearly, the Maya had the capability for investigating and recording a phenomenon
  such as a meteor shower.  The question was, was that astronomical information completely
  lost or merely embedded in the extant inscriptions?

The ancient Maya area covers the northern latitudes from about 14$^{\circ}$
to 21.5$^{\circ}$N and western longitudes from about 87$^{\circ}$ to
93$^{\circ}$W, including the modern Central American countries of eastern
Mexico, Guatemala, Belize, El Salvador and western Honduras.  Although the
Eta Aquariids are considered primarily a southern latitude shower, the
radiant would have been visible to the Maya in the east for more than three
hours before morning twilight.  Without any known recorded radiant
information, the authors' approach in this paper is to compare the date and
time of any computed outbursts to events recorded on or near that date\footnote{All
dates and times throughout this treatise are in the Julian Calendar, UT.  To convert
UT to local (Mexican) time, subtract 6 hours.  When converting from the Christian 
calendar to the Maya calendar, the authors used the correlation constant of
584286 (see \cite{ms12}, \cite{kennetetal13}).  The correlation constant represents 
the Julian Day Number that corresponds
to the Maya Long Count of 13.0.0.0.0, referred to as the ``Creation Date''.
584286 corresponds to 3114 BC September 9, 12:00 UT, Julian Calendar.  The
day following the creation date would have been written 13.0.0.0.1, 
a Julian Day Number of 584287.  Other commonly accepted correlations are
584283 and 584285.  Conversion from Maya calendar to Julian Day
Numbers accomplished using online software by  \cite{pau16}.  Conversion from
Julian Day Number to Julian Calendar accomplished using software by 
Simulation Curriculum \cite{sn09}.  Solar longitudes and  $\Delta{T}$ 
(Table \ref{table:Model Comparison} only; other tables \citet[p.~78]{meeus00}) 
obtained through  JPL'S Horizons system \citep{giorgini96}
}.  

An ``event'' refers to any recorded information 
as described earlier in this section.  Those events and
associated protagonists and dates are the subject of this paper.  In addition,
results are computed and compared to ancient Eta Aquariid dates in the
Chinese record (Table \ref{table:Historical Outbursts}) and Vaubaillon's
computations \citep[table 5e]{jenniskens06} of possible historic Eta Aquariid
outbursts (Table \ref{table:Model Comparison}).

\subsection{Previous attempts at identifying observations of meteor showers}


\citet{hagar31} wrote that the Mexicans, pre-dating the arrival of the
Spanish, commemorated falling stars called ``Tzontemocque or Falling
 Hairs'' by the celebration of an annual festival called Quecholli. 
He maintained that falling figures shown in various Mexican codices such as
the Borgia \citep{famsi16b} and Vaticanus 3773 \citep{famsi16c}
represent a meteor shower, possibly the Leonids and
another figure the Taurids.  \cite{kohler02} notes that the Aztecs recorded a meteor
in 1489 in the Telleriano-Remensis \citep{famsi16a} 
on page 39V of that codex
\citep[see also][p.~287-290]{taube00}.

By using a one day shift for every 71 years for Earth's axis precession,
\citet[p.~112-3]{trenary87} calculated a possible Leonid shower date
in the Maya corpus within a few days of 709 October 
28\footnote{The date is found on a stone panel known as Lintel 24 at the 
site of Yaxchilan.}, although the precession additionally of Leonid orbits 
themselves suggests the expected date around AD 709 may be 2-3 weeks
earlier than October 28 \citep{ahn05}. 

\citet[p.~98]{kinsman14} calculated that two Perseid meteor shower dates in
AD 933 and 775 are possibly recorded in cognate (similar) 
almanacs found in the codices.  The suspected outburst in 933 falls on the 
same date that China observed a Perseid outburst \cite[p.~203]{zhuang77}
\cite[p.~325]{pxj08}.

Therefore other than sidereal year calculations to produce dates that would yield solar longitudes associated with  applicable showers, prior to our investigation there had been no scientific attempts such as numerical integrations by high-speed computers to correlate any ancient Maya dates with any meteor outbursts from any meteoroid streams.

\subsection{Halley's comet and the Eta Aquariids}

The authors decided to investigate Eta Aquariid outbursts.  One reason is
that the orbit of parent comet 1P/Halley, during and for some time before the
Maya Classic Period, is well constrained,
reliable observations dating back to 240 BC in Chinese records
\citep{kiang72} and 164 BC in Babylonian cuneiform texts \citep{stephenson85}.
\cite{yk81} showed that their computed orbit is valid back to
1404 BC, but that Halley's very close approach by Earth
in that year affected the comet's orbit to the extent that computer
models cannot accurately match it at earlier epochs.
Since our study
depends on the meteoroid particles being ejected at each starting epoch,
knowing each exact time is critical in determing the later position of the
particles at a Maya year of observation.  By correcting their computer
model with actual historic observations of Halley's passage by Earth in
837, 374 and 141, \cite{yk81} produced a model with minimum differences in
computed and observed times of perihelion passage, noting (p.~642) the
extrapolated computed times' likely accuracy to better than a month even as
long ago as 1404 BC.

The Halley meteoroid stream produces the Orionids (IAU meteor shower code
00008 ORI) pre-perihelion at the ascending node and the Eta Aquariids (00031
ETA) post-perihelion at the descending node.  The reason to focus here on ETA
is that 1P/Halley's descending node came closest to Earth's
orbit around AD 500 (the ascending node around 800 BC).  Although meteoroid
orbits over time can precess away from the comet orbit to have
nodal intersections at different epochs -- after all, both ORI and ETA showers
are observable at present -- the authors surmised that the chances for the
strongest outbursts in the first millennium AD due to meteoroids released at
recent revolutions of Halley were best for ETA.

Recent orbital analysis by \cite{sw14}
showed that enhanced ETA activity in 2013 was due to dust trails produced by
Halley $\sim$3 kyr earlier, in 1198 BC and 911 BC.  In principle some
outbursts observed by the Maya could be due to trails from before 1404 BC,
but our current aim is to determine observable ETA outbursts from trails
created since then.


\section{Methodology}

Given a starting epoch when particles are released by the comet and an
``end'' year, we consider whether particles from that starting epoch can
reach Earth intersection in that end year, and if so then at what date and
time.

Particles from each return of the comet soon stretch into a trail owing to
variations in initial orbital period.  Particles undergo planetary
perturbations which are a function of where they are along the trail
\citep{plavec56,plavec57}.  If a part of a trail is perturbed to Earth
intersection an outburst occurs.

Instead of period we adopt \da, the difference between particle and comet
semi-major axis $a$ at ejection time, to parametrize the trail.  Similar
1-parameter techniques to identify orbits that intercept Earth at a later
epoch have been used to successfully model meteor outbursts in many streams
(e.g., \citealp{kr85,lyytinen01,maslov11,ma99,sw10,sw14}).

We search for values of \da\ corresponding to particles passing Earth at
small ``miss distance'' $\dr \equiv r_E - r_D$ (\dr\ is proportional to
orbit--orbit minimum distance and is easier to compute) and for such
particles compute also \fm\ ($|\fm|$ represents the along trail spatial
density of particles) and the calendar date when Earth reaches the particles'
descending nodal longitude (essentially the peak outburst time); further
explanation of these quantities is in \cite{asher00}.

If a particle is ejected tangentially at perihelion with relative
speed $\Delta V_T$, then for 1P/Halley's orbit,
\begin{equation}
\frac{\da}{\Delta V_T} \approx 0.04 \, \mathrm{au/(m/s)}.
\label{eqn-dadvt}
\end{equation}
Releasing particles in the visual meteor size range 0.5 down to 0.1 cm radius
at a density of 1 g/cc (\citealp{bk09} quote 0.9$\pm$0.5 g/cc for Orionids)
in tangential positive and negative directions at each perihelion passage of
Halley requires velocities $\sim$ 34 to 76 m/s, taking the comet radius as 4
km in the \cite{whipple51} model, i.e., ejection speeds up to $\sim$76 m/s
occur for such particles.\footnote{It can be shown that for isotropic
  ejection directions at a given single speed, the expected value of the
  tangential component is half that speed.}  This is equivalent to a
\da\ range of $\pm$3 au from Halley's $a$ of $\sim$18 au
(Equation \ref{eqn-dadvt}).

Solar radiation pressure increases a particle's orbital period.  The
parameter $\beta$, the ratio of radiation pressure and gravity forces,
depends on particle size and density \citep[pp.~13--14]{burns79}, with a
particle of radius 0.1 cm having $\beta \approx 1.0 \times 10^{-3}$.
Equation (2) of \cite{ae02} following \cite{reznikov83} gives
$$\Delta{V}_T \approx (V_q\beta)/2 =
 54.6 \, \mathrm{km/s} \times 10^{-3} / 2 \approx 27 \, \mathrm{m/s}$$
where $V_q$ is Comet Halley's perihelion velocity $\sim$55 km/s.  Since the
effect when $\beta=10^{-3}$ is equivalent to a change in velocity $\Delta V_T
\approx 27$ m/s, then from Equation \ref{eqn-dadvt} this solar radiation
pressure has the same effect on the period as ejecting a particle with $\da
\approx 1.1$.  Therefore to compensate for solar radiation pressure $\sim$1
au should be added in the positive tangential direction.

Thus the integrations were carried out over a range of 16 to 22 au, the upper
part of the range associated more with smaller meteoroids (larger
$\beta$).  Initially there were 400--600 particles with a typical spacing of
0.01 au or slightly larger.  Interesting intervals of \da\ parameter space,
with particles approaching Earth at $|\dr| \leq 0.01$ au, were then expanded,
additional integrations with a typical spacing of 0.00001 au in \da\ aiming
to identify the exact time and date of the outburst.  A trail's density cross
section is strongly peaked towards the center where \dr=0 \citep{ma99};
numerical experiments \citep[cf.][]{asher08} suggest an ETA trail encounter
can still generate significant meteor activity for $|\dr|$ up to a few times
0.001 au.  Further integrations were performed until the particles converged
on a solution identifying \da, \dr\ and $\mathit{f}_\mathcal{M}$ showing an
outburst at a specific time and date.

Orbits for 1P/Halley's perihelion returns were from \citet[table 4]{yk81}
and initial state vectors of eight planets from JPL Horizons
\citep{giorgini96}.  Computations used the RADAU algorithm \citep{everhart85}
implemented in the MERCURY integrator \citep{chambers99}.
The authors verified \citeauthor{sw14}'s \citeyearpar{sw14} predictions for
the ETA outbursts in 2013 with similar integrations, and using the
same technique considered in detail 55 different ``end'' years found in
the data base of the Maya corpus of inscriptions wherein a possible ETA
outburst might have been recorded.


\section{Results}

\subsection{Maya events}

The most common event and one that could easily be planned to
coincide on or near a meteor shower that occurred in our data set
was the royal accession, a king or queen's assuming rulership 
over a polity (``taking the royal throne'').
There were 14  accession events in the time frame under investigation:  \\

967 BC (\textit{U Kokan Chan} from \textit{Palenque})(5.8.17.15.17)\footnote{Maya
Long Count date, typically composed of 5 digits in a modified
vigesimal system where the 3rd digit from the
right counts in units of 360 days. See \cite{pau16}
for instance.  The Long Count keeps track of numbers of days
 in a manner similar to the Julian Day Number.  
 967 BC is likely a mythological date and \textit{U Kokan Chan}
likely a mythological ruler},

484 (\textit{Yajaw Te' K'inich I} from \textit{Caracol})(9.2.9.0.16), 

511 (\textit{Lady of Tikal} from \textit{Tikal})(9.3.16.8.4), 

531 (\textit{K'an I} from \textit{Caracol})(9.4.16.13.3),

553 (\textit{Yajaw Te' K'inich II} from \textit{Caracol})(9.5.19.1.2), 

572 (\textit{Kan Bahlam I} from \textit{Palenque})(9.6.18.5.12),

636 (\textit{Yuknoom Ch'en} from \textit{Calakmul})(9.10.3.5.10),

639 (``Ruler 2'' from \textit{Piedras Negras})(9.10.6.5.9),

640 (``Ruler A'' from \textit{Coba})(9.10.7.5.9), 

662 (accession 2 of \textit{Muwan Jol? Pakal})(9.11.9.11.3),

686 (\textit{Yuknoom Yich'aak K'ahk'} from \textit{Calakmul})(9.12.13.17.7), 

752 (\textit{Bird Jaguar IV} from \textit{Yaxchilan})(9.16.1.0.0),

781 (unknown ruler from \textit{Los Higos})(9.17.10.7.0), 

802 (\textit{Lachan K'awiil Ajaw Bot} from \textit{La Amelia})(9.18.11.12.0).     \\

The data set also included rare events such as:  \\

644 (\textit{jatz'bihtuun}, ``strike the stone road'' at \textit{Naranjo})(9.10.11.6.12),

790 (\textit{jatz'bihtuun}, ``strike the stone road'' at \textit{Naranjo})(9.17.19.9.1),

849 (\textit{u-pataw kab'aj}, ``forms the earth''? at \textit{Caracol})(10.0.19.6.14).  \\ 

\noindent
The ``strike the stone road'' event is unique because there are
 only four such occurrences of this event currently known 
 in the corpus of inscriptions and one incidence of
 the ``forms the earth''? event.

 Outbursts occurring on Period Ending dates\footnote{\label{note-pd-ending}
 Period Ending (\textit{pe}) dates are typically 
 separated by 360 day (\textit{tuun}) intervals 
 where the day or \textit{K'in} position, the most right placed
 digit, and the month or \textit{Winal} position, the second
 most right position, would both be zero.  Higher \textit{pe} dates
 involving 7,200 and 144,000 day intervals in a similar
 scheme are also possible.} would be coincidental.
 Period Endings found in our data set included the years 480 (9.2.5.0.0),
 618 (9.9.5.0.0), 687 (9.12.15.0.0), 752 (9.16.1.0.0) and 756 (9.16.5.0.0).

 Four royal births occurred within the constraints of our data set:
 566 (\textit{Lady B'atz' Ek'} from \textit{Caracol} on 9.6.12.4.16),
 588 (\textit{K'an II} from \textit{Caracol} on 9.7.14.10.8),
 606 (\textit{Hix Chapat} from \textit{Tonina} on 9.8.12.14.17) and
 750 (Ruler 7 from \textit{Piedras Negras} on 9.15.18.16.7).
 A birth occurring near the time of an outburst would likely be coincidental.
   
Altogether we have a data set comprising 55 different years, each with one or
more recorded Maya events (\ref{app-data}).
To investigate outbursts, all years were checked in conjunction with all
possible 1P/Halley starting epochs back to and including the 240 BC return,
46 of them back to 616 BC, 36 to 911 BC and 29 to 1404 BC.  Tables
\ref{table:Maya Outbursts Early} (Early Classic) and \ref{table:Maya
  Outbursts Late} (Late Classic) list the 30 end years having the best
possibility of strong outbursts based on the solution parameters \da,
\dr\ and \fm, and observable time within the Maya's visual range.  Apart from
the computed outburst in 572 due to the 911 BC trail listed in Table
\ref{table:Maya Outbursts Early} there were only 3 somewhat successful
solutions involving trails from earlier than 616 BC, none with good enough
parameters to warrant inclusion in Tables \ref{table:Maya Outbursts Early}
and \ref{table:Maya Outbursts Late}.

Tables \ref{table:Maya Events} and \ref{table:Maya Events--6 days or more
  separation}, which collectively list the same end years as Tables
\ref{table:Maya Outbursts Early} and \ref{table:Maya Outbursts Late}, show
that there may have been two categories of Eta Aquariid outbursts that were
noted by the Maya, one involving outbursts that occurred near the time of a
particular event by plus or minus five days and secondly whereby an outburst
preceded an event by approximately one week up to three weeks.  The second
category presents a more difficult problem of connecting the ETA outburst to
that particular event because of the possibility of an intervening shower --
for instance, an outburst noted \citep[p.~200]{zhuang77},\citep[p.~311]{pxj08} 
on AD 461 April 20 could have been one such shower if 461 was a year of
interest.

The best cases among these 30 identified years during the Maya Classic Period
are discussed individually in Section \ref{sec-stgs}.

\begin{table}
\caption{Possible Eta Aquariid outbursts during the Early Classic Period (AD
  250--600).  Solar longitude in J2000.0, date in Julian calendar, TT
  converted to UT using \citet[p.~78]{meeus00}.
  Tr = Year of Halley perihelion passage.  For negative years, add ($-$1)
  to convert to BC, i.e., ($-$239) + ($-$1) = 240 BC.  Positive/negative \fm\
  is mean anomaly $\mathcal{M}$ at end date decreasing/increasing function of
  \da.  Final quoted decimals are of no significance, accuracy being limited
  by comet input data and by knowledge of the meteoroid ejection model, but
  are retained to enable reproducibility of results if the same data and
  model are used.  Visibility considers whether the computed peak time is
  within the range of possible visual observation (after radiant rise and at
  least half an hour before sunrise); there may be increased activity for up
  to a few hours around this.
}
\centering

\begin{tabular}{@{} c c c r c c c c @{}}
\hline\hline

Yr & $\lambda_\odot$ &Date Time& Tr\0 & \da\ (au) & \dr\ (au) & \fm      & Visibility  \\
\hline

328 & 42.727 & Apr \09 04:19 & $-$239 &   +0.932 & $-$0.00057 & $-$0.009 & $-$3h 41m \\
    & 42.682 & Apr \09 03:12 & $-$239 &   +0.931 & $-$0.00099 &   +0.003 & $-$4h 48m \\
480 & 43.983 & Apr  11 10:54 &  $-$86 &   +1.089 & $-$0.00545 & $-$0.019 & vis rng \\
484 & 42.045 & Apr \09 11:28 &    218 &   +1.730 &   +0.00280 &   +0.162 & +0h 13m \\
511 & 42.879 & Apr  11 06:18 &    141 &   +2.578 &   +0.00273 & $-$0.010 & $-$1h 47m \\
531 & 41.962 & Apr  10 10:39 &    451 &   +0.075 &   +0.00081 &   +1.008 & vis rng \\
    & 41.899 & Apr  10 09:05 &    295 &   +0.048 & $-$0.00155 &   +0.666 & vis rng \\
    & 41.884 & Apr  10 08:42 &    374 &   +0.052 & $-$0.00089 &   +0.759 & vis rng \\
556 & 41.332 & Apr \09 04:50 &    374 &   +2.039 &   +0.00215 &   +0.464 & $-$3h 14m \\
562 & 43.011 & Apr  11 11:21 & $-$163 &   +1.967 &   +0.00022 &   +0.001 & +0h 1m \\
    & 42.991 & Apr  11 10:51 & $-$163 &   +1.967 & $-$0.00022 & $-$0.002 & vis rng \\
    & 42.958 & Apr  11 10:02 & $-$163 &   +1.967 & $-$0.00096 &   +0.013 & vis rng \\
    & 42.951 & Apr  11 09:51 & $-$163 &   +1.967 & $-$0.00111 & $-$0.004 & vis rng \\
    & 41.950 & Apr  10 09:00 & $-$239 &   +3.368 &   +0.00347 & $-$0.015 & vis rng \\
    & 41.947 & Apr  10 08:57 & $-$239 &   +3.367 &   +0.00365 &   +0.008 & vis rng \\
    & 41.907 & Apr  10 07:57 & $-$239 &   +3.376 &   +0.00252 &   +0.012 & $-$0h 12m \\
    & 41.901 & Apr  10 07:47 & $-$239 &   +3.372 &   +0.00265 &   +0.011 & $-$0h 21m \\
    & 41.898 & Apr  10 07:44 & $-$239 &   +3.375 &   +0.00253 &   +0.007 & $-$0h 25m \\
    & 41.895 & Apr  10 07:39 & $-$239 &   +3.372 &   +0.00264 & $-$0.006 & $-$0h 30m \\
566 & 41.963 & Apr  10 09:59 & $-$239 &   +2.098 & $-$0.00131 & $-$0.074 & vis rng \\
    & 41.806 & Apr  10 06:06 & $-$239 &   +2.111 & $-$0.00412 &   +0.069 & $-$2h 3m \\
572 & 42.348 & Apr  10 08:24 & $-$910 & $-$1.414 & $-$0.00167 &   +0.002 & vis rng \\
    & 42.307 & Apr  10 07:23 & $-$910 & $-$1.413 & $-$0.00234 & $-$0.007 & $-$0h 42m \\
    & 42.302 & Apr  10 07:15 & $-$910 & $-$1.412 & $-$0.00236 & $-$0.001 & $-$0h 50m \\
    & 42.274 & Apr  10 06:34 & $-$910 & $-$1.412 & $-$0.00238 & $-$0.006 & $-$1h 31m \\
588 & 43.311 & Apr  11 10:47 &  $-$11 & $-$0.096 &   +0.00506 & $-$0.017 & vis rng \\

\hline
\end{tabular}

\label{table:Maya Outbursts Early}
\end{table}

\begin{table}
\caption{Possible Eta Aquariid outbursts during the Late Classic Period (AD
  600--909).
}
\centering

\begin{tabular}{@{} c c c r c c c c @{}}
\hline\hline

Yr & $\lambda_\odot$ &Date Time& Tr\0 & \da\ (au) & \dr\ (au) & \fm      & Visibility  \\
\hline

614 & 44.320 & Apr  13 03:57 &  $-$86 &   +0.363 &   +0.00467 & $-$0.009 & $-$3h 55m \\
    & 43.624 & Apr  12 10:40 &    530 &   +1.137 &   +0.00575 &   +1.090    & vis rng    \\
    & 43.346 & Apr  12 03:46 & $-$465 & $-$1.189 &   +0.00360 & $-$0.051 & $-$4h 10m \\
    & 43.341 & Apr  12 03:39 & $-$465 & $-$1.192 &   +0.00341 & $-$0.306 & $-$4h 17m \\
618 & 41.644 & Apr  10 10:02 & $-$390 &   +0.670 &   +0.00138 & $-$0.006 & vis rng \\
    & 41.574 & Apr  10 08:18 & $-$390 &   +0.694 & $-$0.00239 &   +0.088 & vis rng \\
636 & 42.712 & Apr  11 03:22 &    530 &   +4.531 & $-$0.00117 &   +1.282 & $-$4h 40m \\
639 & 43.923 & Apr  13 03:58 & $-$239 &   +1.979 & $-$0.00335 & $-$0.027 & $-$3h 54m \\
644 & 43.900 & Apr  12 09:54 & $-$86   &   +2.420 &   +0.00549  & $-$0.016 & vis rng   \\
       & 42.981 & Apr  11 11:05 &    374  &   +1.945   & $-$0.00008 & $-$0.003 & vis rng  \\
662 & 42.294 & Apr  11 09:02 & $-$465 &   +2.635 &   +0.00579 & $-$0.037 & vis rng \\
    & 42.279 & Apr  11 08:40 & $-$465 &   +2.637 &   +0.00557 &   +0.020 & vis rng \\
663 & 43.929 & Apr  13 07:47 & $-$465 &   +0.262 & $-$0.00333 & $-$0.092 & $-$0h 10m \\
    & 43.896 & Apr  13 06:57 & $-$465 &   +0.261 & $-$0.00374 &   +0.018 & $-$1h 0m \\
675 & 44.915 & Apr 14 10:05 & $-$239 & $-$0.465 & +0.00341 &   +0.008 & vis rng  \\
       & 44.781 & Apr  14 06:48 & $-$239 & $-$0.460 &   +0.00193 &   +0.018 & $-$1h 8m \\
687 & 43.677 & Apr  13 05:16 &    218 &   +3.369 &   +0.00001 &   +0.002 & $-$2h 44m \\
    & 40.708 & Apr  10 03:37 & $-$163 & $-$1.302 &   +0.00024 &   +0.120 & $-$4h 23m \\
691 & 43.691 & Apr  13 06:05 & $-$314 &   +0.534 & $-$0.00181 & $-$0.002 & $-$1h 49m \\
716 & 43.496 & Apr  12 10:56 &  141     &   +2.714 & +0.00071   &  +0.042   & vis rng   \\
       & 43.487 & Apr  12 10:42 &  141     &   +2.715 & +0.00048   &  $-$0.016 & vis rng  \\
       & 43.429 & Apr  12 09:16 &  141     &   +2.691 & +0.00251   &  $-$0.577 & vis rng  \\
721 & 43.139 & Apr  12 08:59 &  $-$86 &   +3.794 &   +0.00362 & $-$0.005 & vis rng \\
    & 43.128 & Apr  12 08:42 &  $-$86 &   +3.794 &   +0.00344 &   +0.007 & vis rng \\
750 & 42.782 & Apr  12 10:35 & $-$465 &   +2.578 & $-$0.00306 &   +0.004 & vis rng \\
752 & 42.125 & Apr  11 06:37 &    141 &   +1.616 & $-$0.00435 & $-$0.088 & $-$1h 23m \\
    & 42.120 & Apr  11 06:30 &    141 &   +1.610 & $-$0.00431 &   +0.427 & $-$1h 30m \\
    & 42.115 & Apr  11 06:22 &    141 &   +1.599 & $-$0.00411 & $-$0.678 & $-$1h 38m \\
756 & 42.165 & Apr  11 08:10 &    218 &   +1.785 & $-$0.00521 & $-$0.094 & vis rng \\
    & 41.315 & Apr  10 11:05 &    218 &   +3.901 & $-$0.00019 & $-$0.015 & vis rng \\
781 & 45.619 & Apr 15 07:53 & $-$239 & +1.974 & $-$0.00042 & +0.033 & vis rng \\
790 & 41.601 & Apr  11 11:33 &    218 & $-$1.304 &   +0.00075 &   +0.211 & +0h 21m \\
    & 41.598 & Apr  11 11:29 &    218 & $-$1.304 &   +0.00071 & $-$0.137 & +0h 17m \\
802 & 40.388 & Apr  10 07:17 &    218 &   +0.482 & $-$0.00148 &   +0.002 & $-$1h 1m \\
820 & 42.863 & Apr  12 11:18 &  $-$86 &   +3.795 &   +0.00018 &   +0.002 & vis rng \\
849 & 44.303 & Apr  14 09:36 & $-$465 &   +0.638 & $-$0.00027 & $-$0.022 & vis rng \\
    & 44.279 & Apr  14 09:00 & $-$465 &   +0.638 & $-$0.00040 &   +0.022 & vis rng \\

\hline
\end{tabular}

\label{table:Maya Outbursts Late}
\end{table}

\begin{table}
\caption{Outburst within $\pm$5 Days of Event.
Moon = moon age in days, r or s = rise or set, unk = unknown,
acc = accession, pe = Period Ending, \textit{pat} = \textit{pat-kab} = ``to form the Earth''.  Sites:
WAX = Waxactun, CRC = Caracol, PAL = Palenque, ALS = Altar de Los Sacrificios, PNG = Piedras Negras, 
MRL = Moral-Reforma, TZE = Tzendales, NAR = Naranjo, AML = La Amelia, HIG = Los Higos.
Diff = number of days different.
}
\centering

\begin{tabular}{c c r c c c c r}
\hline\hline

Yr & Outburst & Moon & r or s & Event & Date & Site & Diff \\
\hline

328 & Apr \09 & 13.4 & 10:58s & unk    & Apr  11 & WAX & +2 \\
484 & Apr \09 & 27.3 & 10:08r & acc    & Apr  13 & CRC & +4 \\
531 & Apr  10 &  7.8 & 06:54s & acc    & Apr  14 & CRC & +4 \\
556 & Apr \09 & 13.5 & 10:52s & axe    & Apr  10 & CRC & +1 \\
572 & Apr  10 & 11.1 & 09:44s & acc    & Apr \07 & PAL & $-$3 \\
614 & Apr  13 & 28.3 & 10:42r & tomb   & Apr  12 & CRC & $-$1 \\
    & Apr  12 & 27.4 & 10:07r & tomb   & Apr  12 & CRC & 0 \\
618 & Apr  10 & 10.0 & 08:22s & pe     & Apr  14 & ALS & +4 \\
639 & Apr  13 &  3.8 & 03:27s & acc    & Apr  13 & PNG & 0 \\
644 & Apr  12 & 29.4 & 11:59r & strike & Apr \09  & NAR  & $-$3 \\
       & Apr  11 & 28.2 & 11:16r & strike & Apr \09    & NAR & $-$2 \\
662 & Apr  11 & 16.5 & 13:21s & acc    & Apr \06 & MRL & $-$5 \\
687 & Apr  13 & 24.8 & 08:45r & pe     & Apr  12 & PNG & $-$1 \\
    & Apr  10 & 21.6 & 06:15r & pe     & Apr  12 & PNG & +2 \\
691 & Apr  13 &  9.4 & 08:00s & tomb   & Apr  12 & TZE & $-$1 \\
750 & Apr  12 &  1.8 & 12:37r & birth  & Apr \08 & PNG & $-$4 \\
756 & Apr  11 &  6.2 & 05:49s & pe     & Apr \09 & PNG & $-$2 \\
    & Apr  10 &  5.4 & 05:01s & pe     & Apr \09 & PNG & $-$1 \\
781 & Apr 15 & 17.4 & 13:29s & acc & Apr 18 & HIG & +3 \\
790 & Apr  11 & 22.5 & 06:49r & strike & Apr  12 & NAR & +1 \\
802 & Apr  10 &  4.3 & 15:19r & acc    & Apr \08 & AML & $-$2 \\
820 & Apr  12 & 25.1 & 08:38r & tomb?  & Apr  13 & CRC & +1 \\
849 & Apr  14 & 17.5 & 13:47s & \textit{pat} & Apr  15 & CRC & +1 \\ 

\hline
\end{tabular}

\label{table:Maya Events}
\end{table}

\begin{table}[t]
\caption{Outburst Preceding or Following Event by Seven Days or More.
Events: pe = Period Ending, star = Star War (a conquering of one polity over another), 
acc = royal accession, ded = dedicatory event, arr = arrival, 
Sites:  QRG = Quirigua, TIK = Tikal, CRN = La Corona, 
CLK = Calakmul, YAX = Yaxchilan, DPL = Dos Pilas.
}
\centering

\begin{tabular}{c c r c c c c r}
\hline\hline

Yr & Outburst & Moon & r or s & Event & Date & Site & Diff \\
\hline

480 & Apr  11 & 15.6 & 12:20s & pe     & Apr  18 & QRG & +7 \\
511 & Apr  11 & 26.9 & 10:22r & acc    & Apr  20 & TIK & +9 \\
562 & Apr  11 & 21.5 & 05:54r & war   & Apr  30 & CRC & +19 \\
       & Apr  10 & 20.3  & 04:57r & war   & Apr  30 & CRC & +20 \\
566 & Apr  10 &  5.0 & 05:23s & birth  & Apr  23 & CRC & +13 \\
588 & Apr  11 &  8.8 & 07:57s & birth  & Apr  19 & CRC & +8 \\
636 & Apr  11 & 28.4 & 11:01r & acc    & Apr  29 & CLK & +18 \\
663 & Apr  13 & 29.3 & 11:44r & ded    & Apr  23 & CRN & +10 \\
675 & Apr  14 & 13.3 & 10:58s & emerge & Apr  26 & CRN & +12 \\
716 & Apr  12 & 15.5 & 12:29s & war     & Apr 4     & NAR  & $-$8  \\
721 & Apr  12 & 10.0 & 08:50s & arr    & Apr  27 & CRN & +15 \\
752 & Apr  11 & 21.8 & 06:17r & acc    & Apr  30 & YAX & +19 \\

\hline
\end{tabular}

\label{table:Maya Events--6 days or more separation}
\end{table}

\begin{table}
\caption{Data for Historically Observed Eta Aquariid Outbursts (China).
  Observed outbursts compared to integrations. Visible range from radiant
  rise to one half hour prior to sunrise, approx.\ 18:20 to 21:10 UT, but computed outburst time vs. actual visual
  range is calculated from the geographical coordinates of the capital city of the ruling dynasty \cite[p.~468]{pxj08}. 
  See also caption to Fig.~\ref{table:Maya Outbursts Early}.  Year 461 results shown for informational
  purposes only (i.e., historical record described an outburst on April 13 but not as ETA, however either
  the month or day was inscribed in error).
}

\centering

\begin{tabular}{@{} c c l l r c c c c @{}}
\hline\hline

Yr &     & \0$\lambda_\odot$ & Date Time     & Tr\0 & \da\ (au) & \dr\ (au) & \fm      &  \\
\hline\hline

74    & Ob  & 41.3   & Apr \06.0     &        &          &            &          &         \\
  BC  & Int & none   &               &        &          &            &          &          \\
\hline
401   & Ob  & 41.5   & Apr \08.7     &             &              &                      &             &           \\       
         & Int & 41.733 & Apr \08 20:37 & $-$390 &  +0.427 &  $-$0.00508 & +0.032 & vis rng \\
\hline
443   & Ob  & 42.0   & Apr \09.9     &              &                &                      &               & 0, +1d     \\
      & Int & 40.917 & Apr \08 18:50 &    295    & $-$0.890 &   +0.00009    &   +0.653 & vis rng \\
      & Int & 41.929 & Apr \09 19:57 &  $-$163 & +1.604    &  $-$0.00044 & $-$0.006 & vis rng \\
      & Int & 41.950 & Apr \09 20:28 &      66    & $-$0.102 & $-$0.00023 &    +0.001 & vis rng \\
      & Int & 41.959 & Apr \09 20:41 &      66    & $-$0.102 & $-$0.00011 & $-$0.006 & vis rng \\
\hline
461   & Ob  &  46.142 & Apr 13  19:30  &       &               &               &                   &             \\
         & Int   & 41.274 & Apr \08 18:33  & 374 & +1.532  &  $-$0.00228 & +1.134  & vis rng  \\
         & Int   & 42.301 & Apr \09 20.02  & 141 & $-$1.998 & +0.00295   & +0.250  & vis rng  \\

\hline
466   & Ob  & 41.0   & Apr \08.8     &           &               &            &          &           \\
      & Int & 40.817 & Apr \08 13:48 &    295 &   +0.905 & $-$0.00244 &   +0.751 & $-$4h 35m \\
\hline
530   & Ob  & 41.5   & Apr \09.7     &        &          &            &          &           \\
      & Int & 41.690 & Apr \09 21:34 &  $-$465  & +0.075 &   +0.00312 &  +0.204 & +0h 2m \\
\hline
839   & Ob  & 43.2   & Apr  13.7     &        &          &            &          &           \\
      & Int & 43.269 & Apr  13 18:30 &    141 &   +1.736 & $-$0.00085 & $-$0.144 & $-$0h 20m  \\
\hline
905   & Ob  & 43.3   & Apr  13.7     &        &          &            &          &   0, +2d    \\ 
      & Int & 41.432 & Apr  11 18:56 &  $-$11 &   +0.901 &   +0.00231 & $-$0.007 & vis rng \\
     & Int  & 41.432 & Apr 11 18:56  &  $-$11  &   +0.935 &   $-$0.00348 & +0.030 & vis rng \\
    & Int & 43.527 & Apr 13 22:54  &  $-$163 & +1.930 &   +0.00226 & +0.003  & +1h 15m  \\
\hline 
927   & Ob  & 42.7   & Apr  13.7     &        &          &            &          & $-$1d     \\
      & Int & 43.877 & Apr  14 22:57 & $-$314 &   +0.568 & $-$0.00105 & $-$0.005 &   +1h 52m \\
      & Int & 43.894 & Apr 14 23:21 & $-$314  &  +0.570 & $-$0.00093  & +0.005  & +2h 16m  \\
\hline
934   & Ob  & 42.9   & Apr  13.7     &        &          &            &          &           \\
      & Int & 43.290 & Apr  14 03:41 & $-$465 &   +2.909 &   +0.00476 & none     &   +6h 41m \\

\hline\hline
\end{tabular}

\label{table:Historical Outbursts}
\end{table}

\begin{table}[!ht]
\caption{Comparison of this work (K-A) with Vaubaillon's results in
  \citet[table 5e]{jenniskens06}.}

\centering

\begin{tabular}{c c c c r l c c c}
\hline\hline

Yr   &        & \0$\lambda_\odot$   &  Date  Time  &  Tr\0  &  \da\ (au)  &  \dr\ (au)  &  \fm  &  ZHR   \\

\hline\hline

511  &  K-A  & 42.879  &  Apr 11  06:18  & 141   & +2.578        & +0.00273     & $-$0.010  &      \\
        & K-A  &  41.151 &  Apr \09 11:16  &  374  &  $-$1.428   & $-$0.00905  & +0.605    &        \\
       & 5e   &  41.594 &  Apr \09 23:39  &  374  & $-$1.4411 & $-$0.00098     & -.-            &  60   \\
\hline

531  & K-A  &  41.962  &  Apr 10  10:39  & 451  & +0.075     &  +0.00081    & +1.008   &      \\
        & K-A  &  41.899  &  Apr 10  09:05  & 295  & +0.048     & $-$0.00155  & +0.666  &       \\
        & K-A  &  41.884  &  Apr 10  08:42  & 374  & +0.052     & $-$0.00089  & +0.759  &       \\
     & K-A  &  41.464 &  Apr \09 22:07  &  218  & +0.061    & $-$0.00905  & +0.002  &       \\
     & 5e   &  41.935 &  Apr 10  11:10  &  218  &  +0.0472  & $-$0.00109     &  -.-         & 900  \\
\hline

539 & K-A &  42.285 & Apr 10 19:35  & 141   &  +0.412   &  +0.00145   & $-$0.527  &          \\
     & 5e   &  42.388 &  Apr 10  23:27  &  141  &  +0.4818 &  $-$0.00090   &    -.-          & 1200  \\
\hline

543 & K-A &  41.983 & Apr 10  12:57  &  66    & +0.526    & $-$0.00486  &  +0.057  &            \\
     & 5e   &  42.263 &  Apr 10  21:12  &   66  &  +0.5126  &   $-$0.00259  &   -.-         &   20     \\
\hline

550  & K-A  &  40.907 &  Apr \09 05:00  &  451  & +2.918    & $-$0.00842  &  +1.171  &            \\
        & 5e   &  41.342 &  Apr \09 17:06  &  451  &  +2.8680  &   $-$0.00066   &   -.-         &   150   \\
\hline

601 & K-A  &  41.285 &  Apr \09 16:25 &  451  &  $-$0.637   & $-$0.00761  &  +0.434  &        \\
       & 5e   &  41.713 &  Apr 10  04:12  &  451  & $-$0.6420 &   $-$0.00122   &    -.-        &  470  \\
\hline

619  &  K-A  &  41.116  &  Apr 10  03:07  &  374  &  +0.670      & $-$0.00599  &  +0.217  &        \\
        & 5e   &  41.518 &  Apr 10  14:13  &  374  &  +0.6629   &  $-$0.00034  &   -.-         & 190  \\
\hline

641  &  K-A  &  41.529  &  Apr 10  04:40  &  374   &  +1.688    & $-$0.00341  & $-$0.023  &        \\
        &  K-A  &  41.441  &  Apr 10  02:30  &  374   &  +1.720    & $-$0.00557  &  +1.210    &         \\
     & 5e   &  41.827 &  Apr 10  13:07  &  374  &  +1.6986  &   $-$0.00043   &   -.-          &   60  \\
\hline

647  &  K-A  &  41.014 &  Apr 10  04:51  &  451  &  +2.743     & $-$0.00692  &  +1.690    &         \\
        & 5e   &  41.377 &  Apr 10  14:53  &  451  &  +2.7166   &  $-$0.00113   &  -.-            & 370   \\
\hline

650  & K-A  &  40.841 &  Apr \09 19:03  &   66  &  +0.834    & $-$0.00638   & +0.006    &           \\
        & 5e   &  41.139 &  Apr 10  03:28  &   66  &  +0.8581  &  $-$0.00224     &  -.-           &  50     \\
\hline

672  &  K-A &  none   &  none           &  451  &  $-$0.854   &  none         &              &          \\
        & 5e   &  41.843 &  Apr 10  12:17  &  451  & $-$0.871  &  $-$0.00082     &  -.-         &  40      \\
\hline

692   &  K-A  &  43.194 &  Apr 11 23:46  &  530  &  +0.871   & $-$0.00533     &  +0.665  &            \\
         & 5e   &  43.604 &  Apr 12  10:51  &  530  &  +0.8601  &   +0.00024     &  -.-          &  20     \\
\hline

713  &  K-A &  none   &  none           &  141  &  $-$0.820   &  none         &              &          \\
     & 5e   &  43.469 &  Apr 12  16.43  &  141  & $-$0.8347 &  $-$0.00100       &  -.-          &  40    \\
\hline

719  & K-A  &  40.000 &  Apr \09 14:53  &  218  &  +0.468     & $-$0.00429    &  +0.022   &          \\
        & 5e   &  40.263 &  Apr \09 22:17  &  218  &  +0.4707   &  $-$0.00068     &   -.-          &  410  \\
\hline

796   &  K-A & 41.303 &  Apr 10  16:59  &  218  &  +1.306      & $-$0.00434   &  $-$0.130  &         \\
         & 5e   &  41.636 &  Apr 11  01:54  &  218  &  +1.2751   &  $-$0.00084   &    -.-           &  250  \\
\hline

964   &  K-A & 41.888  &  Apr 12  09:18  &  218  &  +0.476    & $-$0.00206    &   +0.082    &          \\  
         & 5e   &  41.973 &  Apr 12  11:44  &  218  &  +0.4755  &  $-$0.00152      &  -.-             &  1100  \\

\hline\hline
\end{tabular}

\label{table:Model Comparison}
\end{table}

\subsection{Historical Eta Aquariid outbursts observed from China}

Our integrations of recorded observations in the Chinese record
during the years parallel to the Maya Classic Period showed
a high correlation to those dates and times and validated our model.
  Table \ref{table:Historical Outbursts}
shows computed results for ancient observations that are all attributed to
China \citep[pp.~199-200]{zhuang77}\citep[p.~134, table 1]{ih58}
\citep[pp.~309-325, 648-659]{pxj08}.  Comet Halley input parameters were the same as for the Maya end
years.

Several integrations, 401, 443, 466, 530, 839, and 905  
correlated directly to the associated dates of
the observed outbursts; 401, 443, 530 and 839 showed computed outburst times
either within or within a few minutes of the visual range
while computed times in
905 and 927 were slightly over an hour outside the visual range.
The 927 return differed by one day from the recorded observation.
There was a very strong 443 return on April 8 one day prior to the reported outburst
followed by another outburst on April 9, the recorded date.  Similarly, in 905
a moderate outburst occurred two days prior to the recorded outburst on April 13.
The historical record for a 461 outburst is not classified as an ETA by
\cite{zhuang77} or \cite{ih58}, however integrations showed strong outbursts
visible from China on both April 8 and 9.  Therefore it is possible that 
there actually was an ETA outburst in 461 where the date registered was in error.

Integrations for the years 466 and 934 showed outbursts outside of the visual range by over 4 hours
and 7 hours respectively, and no result was found for 
the 74 BC outburst so it is possible that those outbursts originated prior to 1404 BC.

\subsection{Comparison with table 5e of \cite{jenniskens06}}

\label{compare 5e}

The authors ran integrations for the trail/year combinations listed in
\citet[table 5e, p.~666]{jenniskens06}.  Results are also included in Table
\ref{table:Model Comparison} from different trails that produced outbursts in
the same end years as listed in Table \ref{table:Maya Outbursts Early}.  The
results with date and times were similar where heavy outbursts were noted,
such as in the years 531, 539 and 964.  Compared to the intense outbursts in
531 from three different trails (Section \ref{531}), the 218 trail would have
produced a very light outburst in our model.  The outburst in 511
computed by the authors and 
possibly noted by the Maya in their inscriptions would have 
been due to a different trail other than was reported in table 5e;
also the 511 outburst as shown in table 5e would not have
been observed in the Maya area as noted by the time of outburst.
The strong outburst in 964
has not been noted in any of the extant inscriptions so far in the Maya record, but 964 is
later than the Classic Period by over 50 years, so this would not be a surprise.
The only other time of outburst noted in table 5e that might have been visible to the
Maya was 692.  The authors' model however shows a time well outside of the visible
range in the Maya area and there was no April date recorded in 692.

\section{Discussion of possible sightings}
\label{sec-stgs}

\subsection{Ordering of outburst intensity}
\label{sec-ord}

Among the 30 years in Tables \ref{table:Maya Outbursts Early} to
\ref{table:Maya Events--6 days or more separation}, stronger outbursts will
be associated with smaller $|\dr|$, higher $|\fm|$, and $\da$ closer to 0 (or
closer to about +1 au for smaller particles).  The likelihood of sightings by
the Maya also depends on the peak time being within or close to the visible
range, and on the phase of the moon if present.  Based on these points the
five most probable ETA displays are (order of descending intensity): \\

531, 566, 618, 663, 849. \\

\noindent
Outbursts less likely though still with a relative high probability of being
observed are (loosely in descending order of likelihood): \\

756, 790, 644, 721, 562, 572, 675, 752, 484, 781, 716, 511 \\

\noindent with others listed in Tables
\ref{table:Maya Outbursts Early}--\ref{table:Maya Events--6 days or more separation}
having somewhat smaller possibility of being observed.

The following possible sightings of outbursts are described in order of their
relative strength or intensity, the strongest being first, then the second
strongest and so on.

\subsection{Extreme outburst in AD 531}
\label{531}

The outburst on 531 April 10, the strongest by far noted by the authors, resulted from
particles released by Halley from three different perihelion passages, AD
295, 374 and 451.  The parameters from each of these trails, low $|\dr|$ and
high $|\fm|$, indicate that any one trail would have produced a very strong
outburst, all three being within the time the radiant would have been visible.
The \fm\ values close to 1 indicate particles much more compressed in the
along trail direction compared to most other cases in Tables \ref{table:Maya
  Outbursts Early} and \ref{table:Maya Outbursts Late}.  The miss distances
$\Delta{r}$ = +0.0008, $-$0.0009 and $-$0.0016 au for the 451, 374 and
295 trails respectively were near optimum for a strong outburst.  With
$\Delta\mathit{a}_0$ so close to zero (Table \ref{table:Maya Outbursts
  Early}), i.e., particle orbits similar to Halley, the particles had not
been ejected very far from the comet indicating heavy and densely-packed
particles that would cause an intense outburst.  The sky was dark since the
moon had set a few hours prior to the radiant rise (Table
\ref{table:Maya Events}) making for even a more
impressive display.  This shower was likely the most intense that the Maya
would have seen during the Classic Period.  A ZHR = 900 was post-dicted for
this same outburst \cite[table 5e]{jenniskens06}, 
also shown in comparison in Table \ref{table:Model Comparison}:
although both models compare favorably in intensity and 
time of outburst, they differ in the responsible trail(s).

    An accession to
the royal throne followed this outburst by 4 days (9.4.16.13.3); 
the likelihood of the connection of an accession
event to this outburst may be strengthened by the fact that 
the inscribed lunar information supplementing the Maya
Long Count indicates a lunar age of 8 days, the actual age of the moon during
the outburst on April 10, not the moon age of 12 days required for the actual
calendar date; whether a scribal error or a notation made
on purpose to indicate the date of an astronomical event is not known at this
time.

\subsection{Outburst in 566 due to 240 BC trail}

From its 240 BC passage Halley produced one relatively moderate outburst 
on the morning of 566 April 10 at about 10:00.  An earlier outburst computed at
about 06:00, $\sim$2 hours prior to radiant visibility was likely not visible.
The visible display at 10:00 had a moderate $|\fm| \sim 0.07$ 
and $|\dr|$ was slightly greater than 0.001 au.  The
moon would not have been a factor since it set a few hours prior to the rise
of the radiant.  Almost two
weeks after this outburst the birth of a princess was
 recorded on 9.6.12.4.16 at the site of Caracol.

\subsection{Outburst in 618 due to 391 BC trail}

Two nearby segments of the 391 BC trail reached Earth on the morning of 618
April 10.  The first outburst peaked at 08:18 and the second at 10:02.  The
first may have been stronger due to \fm=+0.088 versus $|\fm| \sim 0.006$
for the second even though $\Delta\mathit{r}$ = --0.0024 for the
first versus a closer $\Delta\mathit{r}$ = +0.0014 for the second.
A Period Ending fell on 9.9.5.0.0, four days following the outburst.
The age of the moon is inscribed as 11
days which corresponds to within one day of the age of the moon on the
outburst, not the age of the moon that would be required for the inscribed
Long Count (note similar situation for the 531 outburst).

\subsection{Outburst in 663 due to 466 BC trail}

The outburst on the morning of April 13 was from two parts of the trail, the
first occurring about one hour before radiant rise, and
the second at 07:47, a few minutes prior to radiant visibility.  
The other parameters seem to indicate a moderate
outburst, $\dr = -0.0037$ and $-$0.0033, $\Delta\mathit{a}_0$ = +0.26, and
moderate \fm, although the outburst occurring at radiant visibility had
a significantly stronger $|\fm| \sim 0.09$.   The moon was not a factor, 
nearly new and rising slightly after sunrise.  Ten days after the outburst 
there was a house dedication on 9.11.10.12.5 at the site of 
La Corona (CRN).

\subsection{Outburst in 849 due to 466 BC trail}
\label{849}

For the outburst that may have been observed by the Maya on 849 April 14 
a dual intercept was computed for the 466 BC trail
with both solutions close to $\Delta{a}_0 \sim{+0.638}$ au.  Both outbursts
were in the visual range, the first occurring at 09:00 and the second at
09:36, each with a modest $|\fm| \sim 0.02$.  $\Delta{r}$ was very close to
scoring a direct impact, $-$0.0004 au for the first outburst and $-$0.0003 au
for the second.  Although the 13.5 day old moon did not set until about two
hours after sunrise and may have affected viewing somewhat, 
considering the values for all parameters, the dual
outburst likely would have been relatively strong.
 The 849 outburst was significant because the next day on 10.0.19.6.14 a phrase
possibly meaning ``he/she/it forms the earth,'' \textit{u pataw kab'aj}, 
was inscribed on stone monuments Stela 17 and Altar 10 at Caracol \cite[p.~88, 89]{gm04}.  
The phrase seems to occur only once in the hieroglyphic corpus, although the root of the 
verb, \textit{pat} is fairly common.  The possibility of the action described at Caracol 
being related to meteors is intriguing and worthy of further investigation.

\subsection{Outburst in 756 due to 218 trail}

It was possible outbursts occurred both on April 10 and 11.  The outburst on April 10 had a very
low $\Delta\mathit{r}$, $-$0.0002, but the particles were very small,
indicated by $\Delta\mathit{a}_0$ = +3.9 au.
Fortunately the sky would have been dark since the moon had set a few
hours before radiant rise.  The outburst on the morning of the 11th consisted of
medium sized particles,
$\Delta\mathit{a}_0$ = +1.8 au but the only drawback would have been a value of
$\Delta\mathit{r}$ of just over 0.005 au.  The related Maya event was a
Period Ending (9.16.5.0.0) that fell one or two days prior to the
outburst, with several different polities marking the occasion 
with elaborate celebrations\footnote{\label{dance}
  Yaxchilan marked the event by a
  ``dance'' (Lintel 3) and blood letting (Lintel 54) and Quirigua by a
  ``vision event'' \cite[p.~100-104]{looper03} for example.}.

\subsection{Outburst in 790 due to 218 trail}

\label{790}

This possible outburst on April 11
 would have been caused by two adjacent segments of
Halley's trail from AD 218. The numbers are robust, $\Delta\mathit{r} \sim
0.0007$ au, $\Delta\mathit{a}_0 \sim -1.3$ au and $|\fm| \sim 0.2$;
the outburst occurred in the morning twilight and the 22.5 day old moon rose
at 06:49 possibly only hampering viewing conditions slightly.  The following
day (9.17.19.9.1) a ``strike the stone road''   (\textit{jatz'bihtuun}) \citep{stuart07},
\cite[p.~20, 38, 70]{gm04} event was recorded at the site of
Naranjo.

\subsection{Outbursts in 644 due to trails from AD 374 and 87 BC Halley passages}
\label{outburst 644}

Outbursts in 644 occurred under dark skies (new moon rising) 
on both April 11 and April 12 due to
the dust trails from 374 and 87 BC respectively.   Although
$|\fm| = 0.003$ was rather modest on the morning of
April 11, the stream of medium-sized particles impacted the 
Earth in a virtual direct hit as $\Delta\mathit{r} = 0.00008$ au.  
The outburst on the second day,
April 12, may have been lighter since $\Delta\mathit{r} \sim
0.005$ au, although $|\fm| \sim 0.02$ somewhat stronger than
the day before.  The recorded event, again \textit{jatz'bihtuun}, 
``strike the stone road'' was dated 2 days earlier on April 9 (9.10.11.6.12)
and was inscribed on the same stone panel as the 790 event 
(see Section \ref{790}).

\subsection{Outburst in 721 due to 87 BC trail}

The 721 outbursts that occurred on the morning of April 12
 were light but occurred in a dark sky just as the moon was
setting, therefore the display may have been observed by the Maya.
Fifteen days later a woman known as \textit{`Ix Ti' Kan Ajaw}
arrived at the site of La Corona on 9.14.9.9.14.  The question for
investigation might be ``Did the outburst prompt a departure
from some other locale that was a 15 day's walking journey
from La Corona?''

\subsection{Outburst in 562 due to 164 BC and 240 BC trails}

Eta Aquariid activity that may have occurred in 562 on two successive days
could have been due to no less than 6 intercepts of the 240 BC trail
on the morning of April 10 and 4 intercepts of the 164 BC trail
with Earth on the morning of April 11.  Almost all outbursts
were within or very close to the visual time of observation on both days and
nominal computed times of some intercepts were in rapid succession, within
10-15 minutes of each other, enough that those outbursts could have combined
and thus reinforced their intensity.  The last quarter moon may have affected
viewing conditions slightly.  On the 10th, $\Delta\mathit{a}_0$ was around
3.3 au indicating smaller particles and a finer outburst.  The outburst at about
10:00 (04:00 AM local time) on April 11 would have likely been the stronger
of the two days with $|\dr| < 0.001 $ au and $\Delta\mathit{a}_0 < 2.0 $ au.
A war event known as a ``Star War'' followed this probable outburst by
slightly less than three weeks (9.6.8.4.2) so it cannot be said for certain that the two
are connected\footnote{A possible explanation may be that the Star War, which
  usually recorded one polity's defeat of another, began on the day of the
  outburst and finished on the recorded date.  Most of the other Star War
  dates do not correlate directly to solar longitudes of meteor showers that
  would have been visible to the Maya \citep[p.~92]{kinsman14} and thus it cannot be determined at this time
  whether there is a general trend connecting meteor outbursts to Star War
  events.}.  \citet[p.~89]{mg00} note that the defeat of
Tikal from this Star War event ``would change the course of Early Classic
history.''

\subsection{Outburst in 572 due to 911 BC trail}

In 572 there were multiple intercepts from the 911 BC trail on the morning of
April 10.   Two occurred about 45 minutes prior to the rise of the
radiant and although $|\fm|$ was small, the overlapping
nature of the intercepts may have produced a combined overall display if seen.  
All four were in a very small range $\da = -1.412$ to $-$1.414 au, and \dr\ was around
$-$0.002 au in all cases.  The last of the four
intercepts clearly occurred within the visual observation time.  
The accession of a ruler (\textit{Kan B'ahlam I})
occurred at Palenque on April 7 (9.6.18.5.12), 
three days prior to the outburst on April 10.

\subsection{Outburst in 675 due to 240 BC trail}

A trail encounter was computed in AD 675 around an hour prior to radiant rise
and another during the observable time on the morning of April 14; \fm\ was a
modest +0.02 and $\Delta\mathit{r} < +0.002 $ au
for the first likely outburst if seen and slightly weaker for the
second visual display. 
The nearly full moon set at 10:58 but at least $\da \sim -0.5 $ au implies
quite bright meteors.
     The outburst in 675 was possibly noted by the site of La Corona on a carving known as Panel One 12 days later on 9.12.2.15.11 by a departure event.

\subsection{Outburst in 752 due to 141 trail}

There may have been a significant outburst from particles ejected in AD 141 that
appeared on the morning of April 11.  The Earth intercepted one
particular segment of the trail three times in rapid succession, but all
intercepts were about an hour and a half prior to radiant rise.
Although $|\dr|$ was moderate $\sim 0.004$ au,
\fm\ was strong for all three segments, and $\da \approx +1.6$ au indicating
medium-sized particles.  If the display was seen, the moon may have been a
slight factor, 22 days
old and having risen at 06:17.  Although this outburst may be tied to an
accession event 19 days later (9.16.1.0.0), there seems to be
legitimate rationale for the ruler to have waited that long before taking
the throne.\footnote{The king, \textit{Bird Jaguar IV} of the site of
  Yaxchilan, may have waited so that he could assume the throne on a ``round
  number'' or Period Ending date, Maya Long Count 9.16.1.0.0
  \cite[p.~128]{mg00}.}

  \subsection{Outburst in 484 due to 218 trail}
  
  The outburst in 484 occurred around daybreak on the morning 
  of April 9, which may have diminshed its viewing.
  The parameters were moderate $\Delta\mathit{r}$ = 0.003 au,
  $\da \approx +1.7$ au and  $|\fm| \approx 0.2$. 
  Four days later on April 13 (9.2.9.0.16) a royal 
  accession took place at the site of \textit{Caracol}.

\subsection{Outburst in 781 due to 240 BC trail}

The 781 outburst occurred on April 15.  The small-to-medium sized particles
had good encounter parameters (Table \ref{table:Maya Outbursts Late}) though
likely a modest display around the time of radiant rising, being somewhat
affected by the gibbous moon (Table \ref{table:Maya Events}).  A ruler's
accession at the minor site of Los Higos followed 3 days later on
9.17.10.7.0.

\subsection{Outburst in 716 due to 141 trail}
  
  The outburst in 716 on April 12 was caused by three separate sections of the
  141 trail, the first peak at 09:16, the second at 10:42 followed a few 
  minutes later by the third at 10:56.  Although \fm\ was strong ($\sim 0.6$) 
  with the first intercept, the stream was slightly wide of the mark
  where $\dr \approx +0.0025$.  The second two intercepts had
  weaker \fm\ but were closer to direct impact, 
  $\dr \approx +0.0005$ and $\dr \approx +0.0007$.  
  Unfortunately the moon was full and did not set 
  until 12:29, so many of the light particles ($\da \approx +2.7$) 
  may have been washed out.  An attack by the site of Naranjo
  on an unknown opponent is noted to have occurred
  eight days earlier on April 4 (9.14.4.7.5) \citep[p.~II-55]{gm04}.

\subsection{Modest outburst in 511 due to trail from Halley's AD 141 return}

The outburst on April 11, at 06:18, peaking almost 2 hours before radiant
rise, was likely modest if seen, with trail encounter parameters $\dr \approx
+0.0027$, $\da \approx +2.6$ and $|\fm| \approx 0.01$.  The sky would have
been dark with the moon almost new.  Nine days later on April 20 (9.3.16.8.4)
a queen of only six years old assumed the throne\footnote{Normally
  a female would only accede in extreme circumstances, for instance if there
  was no male heir or the failure of the dynasty was imminent; in addition,
  such an installation of a female required elaborate justification
  \citep{martin99}.} at the site of Tikal.
A meteoric display may have provided a suitable back drop 
for the ceremony starring the young \textit{Lady of Tikal}.\footnote{ 
Numerical integrations also showed an outburst occurring the year before
on 510 April 9 at 08:59:  this moderate outburst was
due to particles ejected by the 374 passage of Halley, where
$\Delta\mathit{r}$ = --0.00239, $\Delta\mathit{a}_0$ = --1.535 and
$\mathit{f}_\mathcal{M}$ was fairly strong $\sim 0.6$.  Reflected light from
the moon may have washed out some of the display however since the nearly
full 13.4 day old moon set at 11:27.  How or if this may
have affected the coronation the following year would be
difficult to assess.}

\subsection{Modest outburst in 639 due to 240 BC trail}

This outburst, among our 30 best candidates (Tables \ref{table:Maya Outbursts
  Early}--\ref{table:Maya Events--6 days or more separation}) though not
estimated as one of the strongest, is notable as the dynamics
involves Saturn (Section \ref{sec-mmr}).
The outburst was computed to peak almost 4 hours
before the radiant was visible and so whether the display was seen
depends on its duration; however, the moon had set and if seen the outburst
may have been stronger than it appears strictly from the 1-parameter dust
trail model since there is a significant \da\ range for which particles cross
the ecliptic plane at a very similar time (albeit not exactly at the dust
trail solution time).  Numerical integrations also indicated a solution on
the morning of April 12 within the visual range, but $\Delta\mathit{r}$ was
greater than 0.006 au, so likely this outburst was low level.  A regal
accession was recorded the same day as the April 13 outburst at 
Piedras Negras on 9.10.6.5.9.

\subsection{Mean motion resonances}
\label{sec-mmr}

The natural tendency of gravitational systems to develop synchronicities
among bodies that are close enough to perturb one another \citep[see for
  instance][p.~9-19]{md99} has affected the dynamics of many trails described
above.  Whereas in the absence of resonant perturbations trail particles will
start to scatter considerably after several revolutions, particles trapped in
resonance by planets could remain for thousands of years in a cluster dense
enough to produce an outburst \citep[cf.][]{eb96,abe99,sa13,sekhar16}.

In \textit{mean motion resonance} the mean motions of a meteoroid particle
and planet are in whole number ratio $p:(p+q)$ and orbital periods in the
inverse ratio, neglecting orbital changes in the slowly varying
angles.  By Kepler's 3rd Law, the particle's semi-major axis
$a$ is constrained to remain at a given value, or in practice to oscillate or
\textit{librate} about that value.  If an idealized point -- a
\textit{resonance center} -- moves around an orbit with orbital period
$P_J(p+q)/p$ at all times where $P_J$ is Jupiter's period, then a resonant
particle periodically drifts in front of and behind this point as $a$
librates.  It can be shown that the maximal extent of this libration, front
to back, measured in terms of mean longitude relative to the resonance
center, is $1/(p+q)$ of the orbit.  It follows that there are $(p+q)$
\textit{resonant zones} around the orbit, in any one of which a particle can
librate.  A perturbation to $a$, e.g., from a close approach to a different
planet, can send the particle out of resonance, after which it will drift
beyond the front or back boundary of the resonant zone.
Particles were confirmed trapped in resonance by verifying that the resonance
variable (\citealp[section 4]{peale76}; \citealp[section 3]{greenberg77}),
also called the resonant argument \citep[chapter 8]{md99} librates \citep[see
  also][]{sa14}.

\begin{figure}[t]
\caption{Resonant argument of Jovian 2:13 resonance for 2 particles ejected
  (dot labeled E) in 466 BC; they are imaginary in the sense that their
  evolution before ejection from 1P/Halley is also shown, this prior behavior
  indicating that they were not immediately resonant in 466 BC.  They
  evidently enter the resonance between 466 BC and 100 BC after which they
  rebound between the back (B) and front (F) of their resonant libration as
  the period varies about its average resonant value.  They were separated by
  just 0.0001 au in \da\ and are indistinguishable in this plot before 849,
  when one approaches Earth and is perturbed out of the resonance while the
  other misses by 0.1 au and stays resonant.
}
\centering
\vspace*{3mm}
\includegraphics[width=\columnwidth]{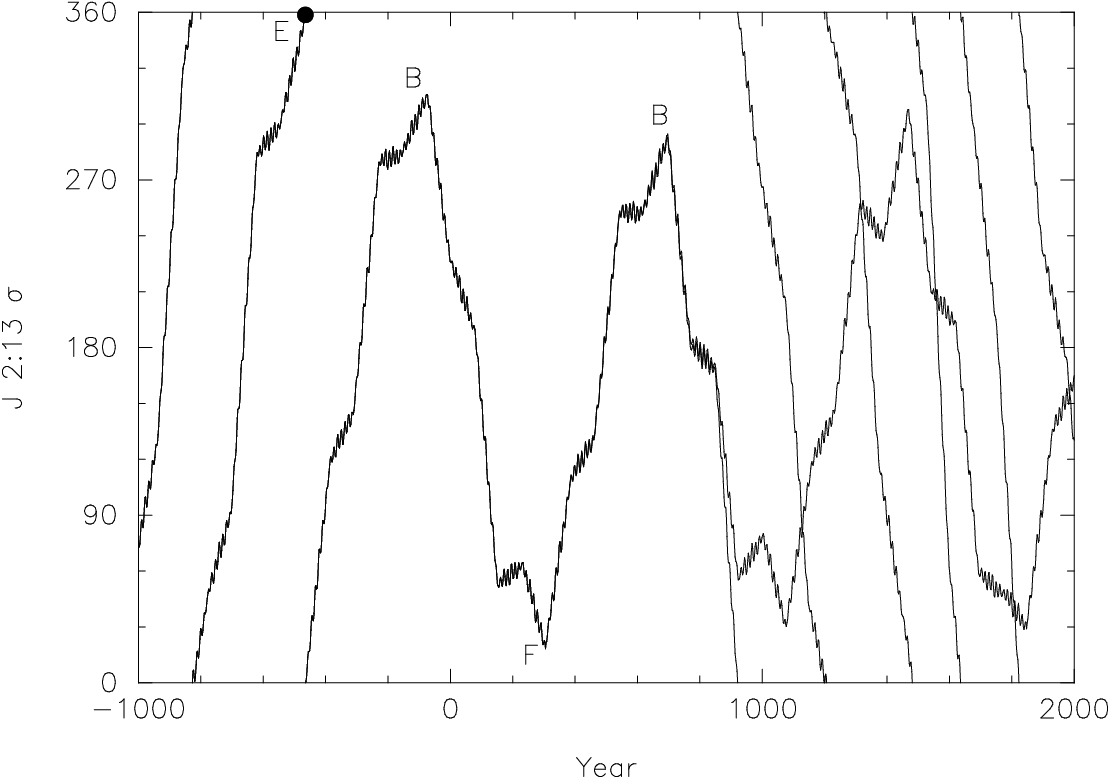}
\label{fig-sigma}
\end{figure}

The 849 outburst (Section \ref{849}) involved the 2:13 Jovian resonance and
particles released at the 466 BC passage of 1P/Halley.  The 2:13 produces 13
zones around the 360$^\circ$ mean longitude of the orbit, any resonant
particle librating within one zone.  Each zone covers $\sim$28$^\circ$ which
is $\pm 14^\circ$ about the respective resonance center.  Figure
\ref{fig-sigma} illustrates resonant trapping, libration vs circulation of
the resonant argument $\sigma$ corresponding to being trapped in that
resonance or not; $\sigma$ effectively amplifies the longitude 13 times so
that the full extent of the resonant zone and the maximum possible peak to
trough libration amplitude are 360$^\circ$.  Here the resonance center is
$\sigma \approx 180^\circ$; Halley in 466 BC (point E in
Fig.~\ref{fig-sigma}) is near the boundary between two adjacent 2:13 zones at
that time.

Selecting a range in \da\ encompassing the solution values, reverse
integrations of 41 ``imaginary'' particles were carried out a few centuries
prior to their release from the comet and then carried forward to AD 2000.
Particles released directly into the resonance would show $\sigma$ libration
before 466 BC.  In fact all these particles were not resonant but soon
afterwards became so.  Although a few particles fell out of resonance (the
particle nearest the solution as a direct result of the 849 Earth encounter:
Fig.~\ref{fig-sigma}), most stayed in the 2:13 zone through 2000.  The action
of the 2:13 covering most of the time frame between 466 BC ejection and
849 Earth encounter ensured a sharp outburst can be produced even after $>$1
kyr.

\begin{table}
\caption{Mean motion resonances, all Jovian except 1:3 Saturnian, causing
  observable outbursts:
  $a_n$ nominal resonance location \citep[sec.\,8.4]{md99};
  $a_0$ osculating semi-major axis at ejection;
  Diff (Tables \ref{table:Maya Events}, \ref{table:Maya Events--6 days or
    more separation}) in days of recorded Maya event.
}
\centering

\begin{tabular}{@{}cccccr@{}}
\hline
Yr  & Trail  & Reson     & $a_n$        & $a_0$          & Diff \\
\hline
562 & 164 BC & 3:23      & 20.22        & 20.047         &  +19 \\
562 & 240 BC & 2:17      & 21.66        & 21.423--21.432 &  +20 \\
566 & 240 BC & 2:15      & 19.93        & 20.154, 20.167 &  +13 \\
572 & 911 BC & 3:17      & 16.53        & 16.377--16.379 & $-$3 \\
618 & 391 BC & 1:7       & 19.03        & 18.628, 18.652 &   +4 \\
639 & 240 BC & 1:3 S     & 19.84        & 20.035         &    0 \\
663 & 466 BC & 3:19      & 17.80        & 18.225, 18.226 &  +10 \\
675 & 240 BC & 1:6, 4:23 & 17.17, 16.69 & 17.596         &  +12 \\
721 &\087 BC & 2:17      & 21.66        & 21.910         &  +15 \\
849 & 466 BC & 2:13      & 18.12        & 18.602         &   +1 \\
\hline
\end{tabular}

\label{table:reson}
\end{table}

Table \ref{table:reson} lists further examples identified by the authors and
indeed four of the best five cases (566, 618, 663, 849: Section
\ref{sec-ord}) are resonant.  The resonances keep particles compact in space
over these time frames, e.g., the particles giving the 572 outburst were
strongly trapped in the 3:17 Jovian resonance for over 1 kyr from ejection
until perturbed by approaching to a few $\times$ 0.01 au of Earth in 234.
When there are only a few centuries between ejection and Earth encounter,
particles may ultimately be resonant but the short time scales render the
resonances irrelevant, with barely time for a full libration cycle.

The 566 and 639 cases (Table \ref{table:reson}) contrast the 2:15 Jovian and
1:3 Saturnian resonances; in the latter the action of another planet than
Jupiter inhibits the dispersion of the particles so that an outburst can
still occur.  The authors verified which resonance operated by plotting the
relevant resonant arguments \citep[cf.][fig.~1]{sa13}.  The segment of the
Halley trail from 240 BC that reached Earth in 639 was in the Saturnian 1:3
during that interval.

Jovian resonance is well known as a cause for historical outbursts,
especially in the Halley stream
\citep[cf.][]{rendtel07,rendtel08,sw07,christou08}.  Comet Halley was in a
1:6 Jovian resonance from 1404 BC to 690 BC, increasing chances that
meteoroids released during this epoch could be trapped in the same resonance,
and was in a 2:13 resonance with Jupiter from 240 BC until AD 1700
\citep{sa14}.

A very strong resonance such as 1:6 can dramatically affect precession rates
which can become much slower.  This explains how Orionid outbursts due to 1:6
meteoroids can occur in the present epoch \citep{sw07}, the precession of
their nodal distance being hugely different from that of Halley whose
ascending node was near 1 au nearly 3 kyr ago.  In many cases the authors
found that the 1:6 substantially slows the precession of the descending
(Eta Aquariid) nodal distance too, potentially making it harder to obtain
1:6 resonant ETA outbursts during the same (Maya) epoch when the comet's
descending node is near 1 au.

\section{Events that preceded ETA solar longitudes by 2--4 days} 
\label{precede eta}

The Maya recorded  a small group of dates that
preceded the likely solar longitude of the ETA's by a few days.  Since it is
fairly certain that the Maya were able to calculate the length of the sidereal year
accurately to at least three decimal places \citep{grofe11,kinsman14},
it would be unusual for the Maya to fall short of a sidereal cycle by 2-4 days.
Therefore it is possible that some rulers were attempting to accede into office
a few days prior to a typical ETA shower.

Assuming that the Maya knew that the peak of the most common meteor showers during the Classic Period occurred on a sidereal year basis, especially the Perseids and Eta Aquariids, the Maya knew that it would have been difficult to synchronize a cycle of a specific shower itself with any of their typical integral number day cycles.  However, they likely realized that the time between peaks of \textit{different} annual showers was an integral number of days; the peak of an ETA (solar longitude $\lambda_\odot \approx 42.0^\circ$) occurred about 266-267
days following the peak of a Perseid  ($\lambda_\odot \approx 139.0^\circ$) the
previous year, or in a minimum day scenario, an ETA would arrive about 262
days after the previous year's Perseid.  A ruler would add
260 days (the length of the sacred Tzolk'in calendar) to the day that the Perseid shower occurred and arrive at a date
that would be no closer than about 2-3 days prior to an expected ETA shower the following year.  If the ruler did not assume office the following year, he would add 365 days or a multiple of that (the length of the \textit{haab'}),
to arrive at the year he expected to take office, to his 260 day calculation.
Table \ref{table:Early Events} shows six examples from
dates\footnote{The Long Count 9.18.6.16.0 listed in Table
\ref{table:Early Events}
is somewhat of an educated guess since only the Tzolk'in date 8 Ajaw is inscribed \citep[see][p.~212, 213]{mg00} and without the additional \textit{haab'} year supplied the selected Long Count is only one of several Long Count options.  However included in the carving of this stone monument (Copan Stela 11) is the phrase ``piercing (by) obsidian,''  and thus a possible connection to meteors; there are only two possibilities of a major shower during the appropriate ruler's reign, the other being an ETA shower.  Dates that are used in
Table \ref{table:Early Events} are just possible dates from the corpus of inscriptions.  The Maya could have easily used other Perseid dates that have not been found in any extant inscriptions.} that are already recorded in the inscriptions wherein the rulers might have applied this simple rule\footnote{The use of combining two cycles is not new, as \cite{powell97} has investigated the number 949 days = 584 days (Venus synodic) + 365 days (\textit{haab'}) and the 819 day cycle in relation to (3)(399) days = 819 + 378, where 378 = Saturn synodic cycle and 399 = Jupiter synodic period.}.
Future research numerically integrating Perseids could shed more light on the Maya's knowledge of the Perseid meteor shower. 

Figure \ref{fig-scatter} in Section \ref{sec-concl}
shows how the accession events from
Table \ref{table:Early Events} are grouped in solar
longitude prior to the most probable outbursts.

 \begin{table}

\caption{Maya Events Occurring prior to Typical Eta Aquariid Showers} Yr (1st column) = year of recorded event.  Ev = event.  ``Long Count'' = the date recorded for that particular event.  ``Perseid'' = the base Long Count that the Maya might have used for computation.  These Long Count dates are already recorded in the inscriptions and also are consistent with solar longitudes of the Perseids.  Yr = year of base computation.  (260) + n(365) = number that the Maya would have added to the base LC (Long Count) to arrive at a calculated pre-ETA date.   260 = 13.0, 365 = 1.0.5 in Maya notation.  n = number of years.   Err = no. of days (d) difference between calculated LC and actual event LC.

\centering

\begin{tabular}{@{} c c c c c c c c c @{}}

\hline\hline

Yr & Ev & Long Count & $\lambda_\odot$ & Perseid & Yr & (260) +    & LC (calc) & err. \\
   &    &  (event)   &     (event)   & (base LC) & (base) & n(365) &           & (d)  \\

\hline

572 &  acc & 9.6.18.5.12 & 39.84 & 9.6.16.10.7 & 570 & (1)(1.0.5) & 9.6.18.5.12 & 0 \\ 
662 & acc  & 9.11.9.11.3  & 37.82 & 9.11.5.15.9.& 658 & (3)(1.0.5) & 9.11.9.11.4 & 1  \\
686 & acc & 9.12.13.17.7 & 35.73 & 9.11.16.0.1 & 668 & (17)(1.0.5) & 9.12.13.17.6 & 1  \\
738 & fire & 9.15.6.13.1   & 36.38 & 9.15.5.0.0   & 736 & (1)(1.0.5) & 9.15.6.13.5   & 4  \\
802 & acc & 9.18.11.12.0 & 38.88 & 9.18.6.16.0 & 797 & (4)(1.0.5) & 9.18.11.12.0 & 0  \\ 
808 & fire & 9.18.17.13.10 & 37.39 & 9.18.6.16.0 & 797 & (10)(1.0.5) & 9.18.17.13.10 & 0 \\

\hline

\end{tabular}

\label{table:Early Events}

\end{table}

\section{Conclusions and Discussion}
\label{sec-concl}

\subsection{Overall conclusion and significant events}

The overall conclusion is that in all probability the Maya kept track of and observed
Eta Aquariid meteor showers and outbursts.

Significantly, the likely most massive display during the Classic Period, the 
outburst of 531, apparently was not missed in the Maya record.
On a moonless night, three very strong, overlapping,
barrages of meteors from the most recent passages 
of Halley impacted Earth within a two hour period followed four days later by
an important Maya royal accession ceremony (\textit{K'an I} on 9.4.16.13.3 at \textit{Caracol}).  

Two \textit{jatz'bihtuun}, ``strike the stone road'' events seem to 
have recorded the observation of an ETA outburst, 
one in 644 and another in 790 (9.10.11.6.12 and 9.17.19.9.1 respectively),
inscribed on the same monument from the site of Naranjo, Guatemala. 
There are only two other known records of this event in the
hieroglyphic corpus, each possibly recording a
numbered shower \citep[pp.~91-92, figure 4]{kinsman14}\citep[pp.~601, 608]{jenniskens06}
or in one case, ``lost'' shower ``D'' \cite[p.~136, table 1]{ih58}.

The weather may have been a factor in a few cases but it is doubtful for instance that 
cloud cover would have prevented the entire Maya population from observing a shower or outburst from every
location in the entire Maya area.  It seems a safe assumption that at least
one site would have had an unobstructed view of the heavens at any time during
the year.

\subsection{ Events with regard to most probable outbursts}
\label{outbursts events}

The likely most intense outbursts computed are paired with recorded Maya
events as follows.
The five most probable ETA outbursts (cf.\ Section \ref{sec-ord}) are: \\

531 (royal accession)(Caracol)

566 (royal birth)(Caracol)

618 (Period Ending)(Altar de Los Sacrificios)

663 (house dedication)(La Corona)

849 (``forms the earth''?)(Caracol).\\
 
And the next ten are: \\

756 (Period Ending with special ceremonies)(multiple sites) 

790 (``strike the stone road'')(Naranjo)

644 (``strike the stone road'')(Naranjo)

721 (arrival)(La Corona)

562 (``Star war'')(Caracol)

572 (royal accession)(Palenque)

675 (departure [emerge?])(La Corona)

752 (royal accession)(Yaxchilan)

484 (royal accession)(Caracol)

781 (royal accession)(Los Higos).  \\
  
Outbursts in 716 (war event, Naranjo), 511 (royal accession, Tikal) and 588 (royal birth, Caracol) would have had
some chance of being observed, and still others listed in Tables
\ref{table:Maya Outbursts Early}--\ref{table:Maya Events--6 days or more separation}
had some albeit small possibility.

\begin{figure}[t]
\caption{Distribution of Most Probable Outbursts in Relation to April
  Accessions, Primordial 3298 BC Event
  (12.10.12.14.18)\cite[p.~68-77]{stuart05}, 967 BC Accession, and Comet
  Halley 374 Perihelion Passage.  Solid triangles with Maya site identifiers
  mark the most probable outbursts determined in this treatise.  Horizontal
  scale is J2000.0 $\lambda_\odot$.  Separation of points on vertical scale
  is for ease of reading only.  } \centering
\vspace*{3mm}
\includegraphics[width=\columnwidth]{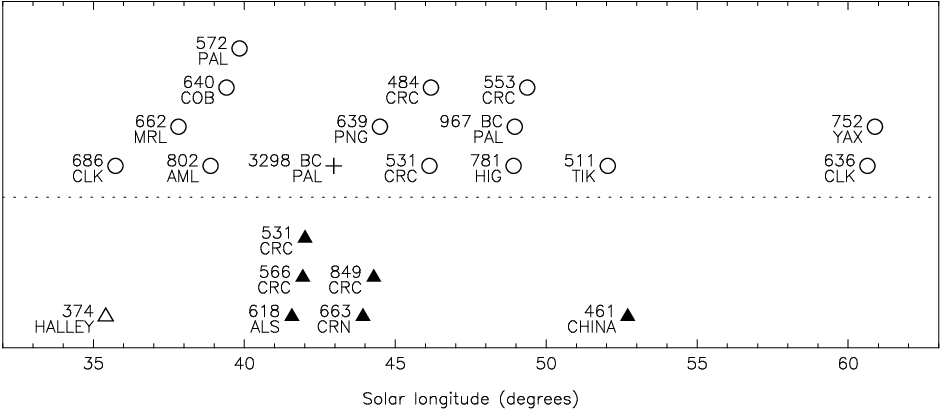}
\label{fig-scatter}
\end{figure}

\begin{figure}
\caption{Distribution of accessions during the 30 day period covering the
end of December and most of January (Maya mid-Classic Period) 
compared to historically observed outbursts.  Horizontal scale 
is J2000.0 $\lambda_\odot$.  Vertical separation for ease of reading only.
}
\centering
\vspace*{3mm}
\includegraphics[width=\columnwidth]{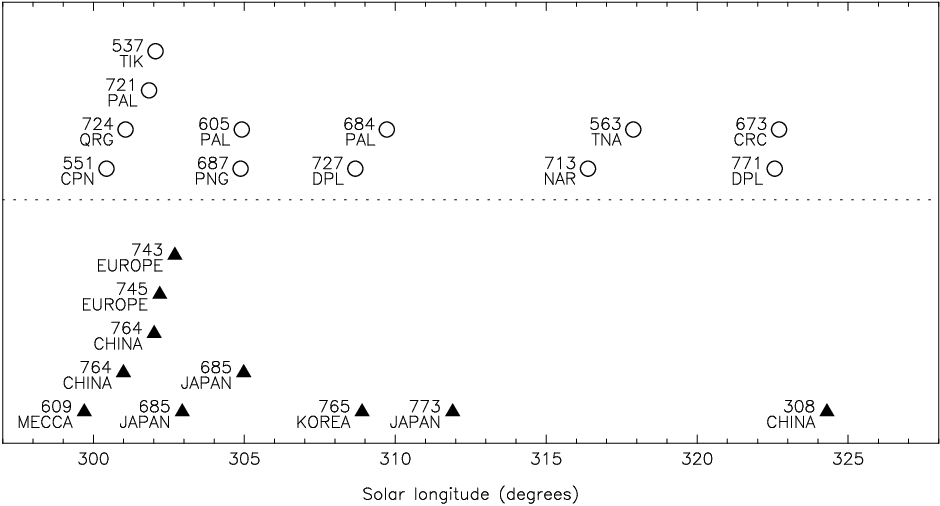}
\label{fig-scatterJan}
\end{figure}

\subsection{Outbursts and ETA's with regard to royal accessions}

Figure \ref{fig-scatter} shows the distribution of the 14 accession events
that occurred  during April, a range of about 
30 days or $30^\circ$ in solar longitude, throughout the Classic Period.  In
a random distribution, there would be slightly less than one accession every
two days; the figure, however, shows at least three different groups, with a
distinct gap between the early event group and the accessions following the
3298 BC primordial event and the most probable outbursts. 

Out of about 70\footnote{A count of individual events in the 
Maya corpus by the authors otherwise not noted in this study.} 
events that occurred in the month of April, the 14 accessions
constitute about 20\%.
The 14 occurring in April of all years out of approximately 
90 recorded royal accessions that occurred during the Classic Period 
represent a disproportionate amount of accessions for one
month.  In another 30 day period approximating the month
of January covering numbered shower 32 
 \citep[pp.~91-92, figure 4]{kinsman14}\citep[p.~610]{jenniskens06},
the authors note up to 12 accessions that could be related to these showers
(see Figure \ref{fig-scatterJan}). 
  The binomial probability 
for 26 or more out of 90 accessions to occur in only two specified months of
the year ($p$=1/6) is 0.0027 (or if multiplied by 66, the number of possible
pairs of months, is still below 2\%).  Therefore it is highly unlikely that
the distribution of these 26 accessions within those two months is random.

There were 30 different years during the Maya Classic Period for which 
integrations revealed that ETA outburst activity may have occurred.
Approximately 18 of those exhibited an especially higher likelihood of actual 
outbursts being observed in the Maya area.   Six of the possible 18 
outbursts occurred near the time of an accession event; those years were: \\

484 (\textit{Yajaw Te' K'inich} from \textit{Caracol}), 

511 (\textit{Lady of Tikal} from \textit{Tikal}), 

531 (\textit{K'an I} from \textit{Caracol}), 

572 (\textit{Kan Bahlam I} from \textit{Palenque}), 

752 (\textit{Bird Jaguar IV} from \textit{Yaxchilan})

781 (unknown ruler from \textit{Los Higos})   \\

The years 511 and 531 are also analyzed by Vaubaillon in
\citet[table 5e]{jenniskens06}, as we described in Section \ref{compare 5e}.
These years constitute pairings of recorded accession events with
computed meteor outbursts.  In the following statistical calculation we
impose the condition that the pair should match within $\pm4$ days which
disqualifies 511 (accession followed outburst by 9 days; Table
\ref{table:Maya Events--6 days or more separation}).

The year 572  may have been the first year that a ruler attempted to forecast an ETA.  Whether expecting an outburst or
simply the annual shower is not known, but several other accessions may have attempted similar
predictions as shown in Table \ref{table:Early Events}.  If this assumption is true, 
it would connect another three (four if 572 is included in the early group
in Figure \ref{fig-scatter}
and not the outburst group) accession events to the ETA's,
bringing the total to seven out of 12 accessions connected to the ETA's.

Considering both types of accessions, i.e., those following outbursts and
those occurring prior to typical ETA solar longitudes as described in Section
\ref{precede eta}, both within a $\pm$4 day period
during the month\footnote{Qualifying
  years for outbursts: 531, 484, 781.  Qualifying years for early accessions:
  572, 662, 686, 802.  Disqualified due to out of $\pm$4 day criterion: 511.
  No outburst or too weak outburst to include: 553, 636, 639 and 640.  Not
  included in $n$: 967 BC, 752.  An accession that occurred in 967 BC, was
  not included in the analysis because there were not enough previous Halley
  perihelion passages available going back to 1404 BC for a statistically
  valid analysis of this event.  For 752 the possibility exists of the ruler
  choosing the accession date for a reason other than an outburst.},
the binomial probability (at least 7 out of $n$=12 with $p$=9/30) implies only
4\% chance of a random occurrence.

The primordial event  \citep[p.~68-77]{stuart05} that the Maya recorded as occurring on 3298 BC March 17
correlates to a solar longitude that is compatible with the ETA's, 43.3$^{\circ}$
(see Figure \ref{fig-scatter}).  For reasons discussed in \citet[pp.~44-45, figure 3]{kinsman15}
the authors believe this primordial event may be linked to the ETA's and 
thus the events discussed herein are not only related to contemporary
ETA's but also the primordial event.  In other cases such as the
site of Palenque it is believed that some rulers related their accessions
to mythological events through sidereal Earth years \citep{kinsman16}.

\subsection{Outbursts related to events other than accessions}

The variety of events paired with the other 12 of the 18 likely strongest
outbursts are: \\

Royal births (566, \textit{Lady Batz' Ek'}; 588, \textit{K'an II})(\textit{Caracol})

Period Endings (618, \textit{Altar de Los Sacrificios}; 756, multiple sites)
 
War (562, ``Star War','' \textit{Caracol} defeats \textit{Tikal};  716, Naranjo attacks unknown opponent)

Departure (675, \textit{Yuknom Yich'ak K'ahk'} from \textit{La Corona})

Arrival (721, \textit{Lady Ti' Kan Ajaw} at \textit{La Corona})  

Building dedication (663, \textit{La Corona})
 
``Forms the earth(?)'' (849, \textit{u-pataw-kab'aj}\footnote{Although 
the root of the verbal phrase \textit{pat} has several different 
meanings and the correct meaning has yet to be
determined \citep[p.~89]{gm04}, the authors believe the stated
meaning fits the context most appropriately.}, \textit{Caracol}) 
 
``Strike the stone road'' (644 and 790, \textit{jatz' bihtuun}, \textit{Naranjo})  \\

\subsection{Spearthrower Owl and Comet Halley}

There may be alternative ways of linking accessions to previous extraordinary events in a sidereal way. 
Heretofore the visible display of 1P/Halley has not been discussed, yet this
comet made its second closest known approach to Earth on 374 April 1 at a
distance of 0.09 au \citep{seargent09,yk81}.  Although the historical
description is somewhat mundane, \cite{seargent09}(p. 40-41) describes its passage as one of the
greatest comets in history:
  
\begin{quote}
  On April 1, it appeared in the south as a broom star, and reached an elongation of 166 degrees from the sun on the third of that month.  This must have been an incredible sight, but (once again) physical description is lacking.  The comet went out of sight sometime during the month of April.  Although there is nothing in the very matter of fact description to suggest it, Halley's Comet at this apparition almost certainly deserves a place among the greatest of the greats.
\end{quote}
  
So, were any of the accessions shown in Figure \ref{fig-scatter} linked to
this incredible sight?  Possibly, of course, since the Maya must have seen
the comet at this passage as the Chinese did \cite[p.~50]{pxj08}\citep{yk81}.
However, there may be a better connection with another accession previously
not mentioned, and that is the accession of a remarkable figure known as
\textit{Spearthrower Owl} on 8.16.17.9.0
\cite[p.~31]{mg00}\cite[p.~13]{martin03}\cite[p.~481-490]{stuart00}. 
Long Count 8.16.17.9.0 corresponds
to 374 May 5, about one month after Halley's 374 passage by Earth.
\textit{Spearthrower Owl}, whose hieroglyph and iconic representations
clearly depict an owl, holds in his hand an \textit{atlatl} with stars attached, an
overt symbol of meteors or ``star darts'' \cite[p.~298]{taube00}.  Amazingly,
\textit{Spearthrower Owl} was probably from the distant non-Maya site of
Teotihuacan, and was responsible for the establishment of a ``New Order'' at
the site of Tikal in 378 \cite[p.~11-15]{martin03}.  Therefore, with the
connection to meteors, and likely comets as well, it may be that
\textit{Spearthrower Owl} based his accession on the passage of Halley in
374, as his accession occurred about one month after the close approach to
Earth, or perhaps a few weeks after its disappearance, not unlike some of the
Maya accessions discussed in this paper shown in Table \ref{table:Early
  Events}.  This connection of \textit{Spearthrower Owl}'s accession to Comet
Halley must nevertheless remain speculative while there is no knowledge of
any observations of comets in the Maya hieroglyphic corpus.

\section*{Acknowledgements}

The authors would like to thank Armagh Observatory and especially Mark
Bailey, Emeritus Director, for the outstanding support during the research
for this paper and Greg Milligan for extraordinary computer technical support. 
J. H. Kinsman especially thanks Prof.\ Bailey and
Dr.\ David Asher for the opportunity to be a guest researcher at the
Observatory during the summer of 2015, during which the initial portions of
the research were conducted.  D.~J.~A. thanks the N. Ireland Department for
Communities for research funding.  The authors are grateful to both reviewers
whose comments led to significant improvements in the paper.

\section{References}

\bibliography{met2016-etaMaya.bib}

\begin{thebibliography}{68}
\expandafter\ifx\csname natexlab\endcsname\relax\def\natexlab#1{#1}\fi
\providecommand{\url}[1]{\texttt{#1}}
\providecommand{\href}[2]{#2}
\providecommand{\path}[1]{#1}
\providecommand{\DOIprefix}{doi:}
\providecommand{\ArXivprefix}{arXiv:}
\providecommand{\URLprefix}{URL: }
\providecommand{\Pubmedprefix}{pmid:}
\providecommand{\doi}[1]{\href{http://dx.doi.org/#1}{\path{#1}}}
\providecommand{\Pubmed}[1]{\href{pmid:#1}{\path{#1}}}
\providecommand{\bibinfo}[2]{#2}
\ifx\xfnm\relax \def\xfnm[#1]{\unskip,\space#1}\fi
\bibitem[{Ahn(2005)}]{ahn05}
\bibinfo{author}{Ahn, S.H.}, \bibinfo{year}{2005}.
\newblock \bibinfo{title}{Meteoric activities during the 11th century}.
\newblock \bibinfo{journal}{Mon.\ Not.\ Roy.\ Astron.\ Soc.}
  \bibinfo{volume}{358}, \bibinfo{pages}{1105--1115}.
\bibitem[{Asher(2000)}]{asher00}
\bibinfo{author}{Asher, D.J.}, \bibinfo{year}{2000}.
\newblock \bibinfo{title}{Leonid dust trail theories}, in:
  \bibinfo{editor}{Arlt, R.} (Ed.), \bibinfo{booktitle}{Proc.\ International
  Meteor Conference, Frasso Sabino 1999}, \bibinfo{publisher}{International
  Meteor Organization}. pp. \bibinfo{pages}{5--21}.
\bibitem[{Asher(2008)}]{asher08}
\bibinfo{author}{Asher, D.J.}, \bibinfo{year}{2008}.
\newblock \bibinfo{title}{Meteor outburst profiles and cometary ejection
  models}.
\newblock \bibinfo{journal}{Earth Moon Plan.} \bibinfo{volume}{102},
  \bibinfo{pages}{27--33}.
\bibitem[{{Asher} et~al.(1999){Asher}, {Bailey} and {Emel'yanenko}}]{abe99}
\bibinfo{author}{{Asher}, D.J.}, \bibinfo{author}{{Bailey}, M.E.},
  \bibinfo{author}{{Emel'yanenko}, V.V.}, \bibinfo{year}{1999}.
\newblock \bibinfo{title}{{Resonant meteoroids from Comet Tempel-Tuttle in
  1333: the cause of the unexpected Leonid outburst in 1998}}.
\newblock \bibinfo{journal}{Mon.\ Not.\ Roy.\ Astron.\ Soc.}
  \bibinfo{volume}{304}, \bibinfo{pages}{L53--L56}.
\bibitem[{{Asher} and {Emel'yanenko}(2002)}]{ae02}
\bibinfo{author}{{Asher}, D.J.}, \bibinfo{author}{{Emel'yanenko}, V.V.},
  \bibinfo{year}{2002}.
\newblock \bibinfo{title}{{The origin of the June Bootid outburst in 1998 and
  determination of cometary ejection velocities}}.
\newblock \bibinfo{journal}{Mon.\ Not.\ Roy.\ Astron.\ Soc.}
  \bibinfo{volume}{331}, \bibinfo{pages}{126--132}.
\bibitem[{Babadzhanov and Kokhirova(2009)}]{bk09}
\bibinfo{author}{Babadzhanov, P.B.}, \bibinfo{author}{Kokhirova, G.I.},
  \bibinfo{year}{2009}.
\newblock \bibinfo{title}{Densities and porosities of meteoroids}.
\newblock \bibinfo{journal}{Astron.\ Astrophys.} \bibinfo{volume}{495},
  \bibinfo{pages}{353--358}.
\bibitem[{Borgia()}]{famsi16b}
\bibinfo{author}{Borgia}, .
\newblock \bibinfo{title}{Codex borgia}.
\newblock \bibinfo{howpublished}{Foundation for the Advancement of Mesoamerican
  Studies, Inc.}
\newblock \bibinfo{note}{Facsimile, Electronic document,
  http://www/famsi/org/research/codices}.
\bibitem[{Burns et~al.(1979)Burns, Lamy and Soter}]{burns79}
\bibinfo{author}{Burns, J.A.}, \bibinfo{author}{Lamy, P.L.},
  \bibinfo{author}{Soter, S.}, \bibinfo{year}{1979}.
\newblock \bibinfo{title}{Radiation forces on small particles in the solar
  system}.
\newblock \bibinfo{journal}{Icarus} \bibinfo{volume}{40},
  \bibinfo{pages}{1--48}.
\bibitem[{Chambers(1999)}]{chambers99}
\bibinfo{author}{Chambers, J.E.}, \bibinfo{year}{1999}.
\newblock \bibinfo{title}{A hybrid symplectic integrator that permits close
  encounters between massive bodies}.
\newblock \bibinfo{journal}{Mon.\ Not.\ Roy.\ Astron.\ Soc.}
  \bibinfo{volume}{304}, \bibinfo{pages}{793--799}.
\bibitem[{{Christou} et~al.(2008){Christou}, {Vaubaillon} and
  {Withers}}]{christou08}
\bibinfo{author}{{Christou}, A.A.}, \bibinfo{author}{{Vaubaillon}, J.},
  \bibinfo{author}{{Withers}, P.}, \bibinfo{year}{2008}.
\newblock \bibinfo{title}{{The P/Halley stream: meteor showers on Earth, Venus
  and Mars}}.
\newblock \bibinfo{journal}{Earth Moon Plan.} \bibinfo{volume}{102},
  \bibinfo{pages}{125--131}.
\bibitem[{Corp.(2009)}]{sn09}
\bibinfo{author}{Corp., S.C.}, \bibinfo{year}{2009}.
\newblock \bibinfo{title}{Starry night pro plus version 6.4.3}.
\newblock \bibinfo{howpublished}{Software}.
\bibitem[{Emel'yanenko and Bailey(1996)}]{eb96}
\bibinfo{author}{Emel'yanenko, V.V.}, \bibinfo{author}{Bailey, M.E.},
  \bibinfo{year}{1996}.
\newblock \bibinfo{title}{Regular and stochastic motion of meteoroid streams in
  {H}alley-type orbits}, in: \bibinfo{editor}{Gustafson, B.A.S.},
  \bibinfo{editor}{Hanner, M.S.} (Eds.), \bibinfo{booktitle}{ASP Conf.\ Ser.\
  Vol.\ 104, Physics, Chemistry and Dynamics of Interplanetary Dust},
  \bibinfo{publisher}{Astron.\ Soc.\ Pacif., San Francisco}. pp.
  \bibinfo{pages}{121--124}.
\bibitem[{Everhart(1985)}]{everhart85}
\bibinfo{author}{Everhart, E.}, \bibinfo{year}{1985}.
\newblock \bibinfo{title}{{An efficient integrator that uses Gauss-Radau
  spacings}}, in: \bibinfo{editor}{Carusi, A.}, \bibinfo{editor}{Valsecchi,
  G.B.} (Eds.), \bibinfo{booktitle}{Dynamics of Comets: Their Origin and
  Evolution (Proc.\ IAU Colloq.\ 83; Astrophys.\ Space Sci.\ Libr.\ Vol.\
  115)}, \bibinfo{publisher}{Reidel, Dordrecht}. pp. \bibinfo{pages}{185--202}.
\bibitem[{Giorgini et~al.(1996)Giorgini, Yeomans, Chamberlin, Chodas, Jacobson,
  Keesey, Lieske, Ostro, Standish and Wimberly}]{giorgini96}
\bibinfo{author}{Giorgini, J.D.}, \bibinfo{author}{Yeomans, D.K.},
  \bibinfo{author}{Chamberlin, A.B.}, \bibinfo{author}{Chodas, P.W.},
  \bibinfo{author}{Jacobson, R.A.}, \bibinfo{author}{Keesey, M.S.},
  \bibinfo{author}{Lieske, J.H.}, \bibinfo{author}{Ostro, S.J.},
  \bibinfo{author}{Standish, E.M.}, \bibinfo{author}{Wimberly, R.N.},
  \bibinfo{year}{1996}.
\newblock \bibinfo{title}{{JPL}'s on-line solar system data service}.
\newblock \bibinfo{journal}{Bull.\ Amer.\ Astron.\ Soc.} \bibinfo{volume}{28},
  \bibinfo{pages}{1158}.
\bibitem[{{Greenberg}(1977)}]{greenberg77}
\bibinfo{author}{{Greenberg}, R.}, \bibinfo{year}{1977}.
\newblock \bibinfo{title}{Orbit-orbit resonances in the solar system: Varieties
  and similarities}.
\newblock \bibinfo{journal}{Vistas Astron.} \bibinfo{volume}{21},
  \bibinfo{pages}{209--239}.
\bibitem[{Grofe(2011)}]{grofe11}
\bibinfo{author}{Grofe, M.J.}, \bibinfo{year}{2011}.
\newblock \bibinfo{title}{The sidereal year and the celestial caiman: Measuring
  deep time in {M}aya inscriptions}.
\newblock \bibinfo{journal}{Archaeoastronomy} \bibinfo{volume}{XXIV},
  \bibinfo{pages}{56--101}.
\bibitem[{Grube and Martin(2004)}]{gm04}
\bibinfo{author}{Grube, N.}, \bibinfo{author}{Martin, S.},
  \bibinfo{year}{2004}.
\newblock \bibinfo{title}{Patronage, betrayal, and revenge: Diplomacy and
  politics in the eastern {M}aya lowlands}, in: \bibinfo{booktitle}{Maya
  Hieroglyphic Forum at Texas}, \bibinfo{organization}{The University of Texas
  at Austin}. \bibinfo{publisher}{Maya Workshop Foundation}. pp.
  \bibinfo{pages}{1--95}.
\newblock \bibinfo{note}{Part II}.
\bibitem[{Hagar(1931)}]{hagar31}
\bibinfo{author}{Hagar, S.}, \bibinfo{year}{1931}.
\newblock \bibinfo{title}{The {N}ovember meteors in {M}aya and {M}exican
  tradition}.
\newblock \bibinfo{journal}{Popular Astron.} \bibinfo{volume}{39},
  \bibinfo{pages}{399--401}.
\bibitem[{Imoto and Hasegawa(1958)}]{ih58}
\bibinfo{author}{Imoto, S.}, \bibinfo{author}{Hasegawa, I.},
  \bibinfo{year}{1958}.
\newblock \bibinfo{title}{Historical records of meteor showers in china, korea,
  and japan}.
\newblock \bibinfo{journal}{Smithsonian Contribution to Astrophysics}
  \bibinfo{volume}{2}, \bibinfo{pages}{131--144}.
\bibitem[{Jenniskens(2006)}]{jenniskens06}
\bibinfo{author}{Jenniskens, P.}, \bibinfo{year}{2006}.
\newblock \bibinfo{title}{Meteor Showers and their Parent Comets}.
\newblock \bibinfo{publisher}{Cambridge University Press}.
\bibitem[{Kennett et~al.(2013)Kennett, Hajdas, Culleton, Belmecheri, Martin,
  Neff, Awe, Graham, Freeman, Newsom, Lentz, Anselmetti, Robinson, Marwan,
  Southon, Hodell and Haug}]{kennetetal13}
\bibinfo{author}{Kennett, D.J.}, \bibinfo{author}{Hajdas, I.},
  \bibinfo{author}{Culleton, B.J.}, \bibinfo{author}{Belmecheri, S.},
  \bibinfo{author}{Martin, S.}, \bibinfo{author}{Neff, H.},
  \bibinfo{author}{Awe, J.}, \bibinfo{author}{Graham, H.V.},
  \bibinfo{author}{Freeman, K.H.}, \bibinfo{author}{Newsom, L.},
  \bibinfo{author}{Lentz, D.L.}, \bibinfo{author}{Anselmetti, F.S.},
  \bibinfo{author}{Robinson, M.}, \bibinfo{author}{Marwan, N.},
  \bibinfo{author}{Southon, J.}, \bibinfo{author}{Hodell, D.A.},
  \bibinfo{author}{Haug, G.H.}, \bibinfo{year}{2013}.
\newblock \bibinfo{title}{Correlating the ancient {M}aya and modern {E}uropean
  calendars with high-precision {AMS 14C} dating}.
\newblock \bibinfo{journal}{Scientific Reports} \bibinfo{volume}{3},
  \bibinfo{pages}{1597 EP}.
\bibitem[{Kiang(1972)}]{kiang72}
\bibinfo{author}{Kiang, T.}, \bibinfo{year}{1972}.
\newblock \bibinfo{title}{The past orbit of {H}alley's comet}.
\newblock \bibinfo{journal}{Mem.\ Roy.\ Astron.\ Soc.} \bibinfo{volume}{76},
  \bibinfo{pages}{27--66}.
\bibitem[{Kinsman(2014)}]{kinsman14}
\bibinfo{author}{Kinsman, J.H.}, \bibinfo{year}{2014}.
\newblock \bibinfo{title}{Meteor showers in the ancient {M}aya hieroglyphic
  codices}, in: \bibinfo{editor}{Jopek, T.J.}, \bibinfo{editor}{Rietmeijer,
  F.J.M.}, \bibinfo{editor}{Watanabe, J.}, \bibinfo{editor}{Williams, I.P.}
  (Eds.), \bibinfo{booktitle}{Proceedings of the Meteoroids 2013 Conference,
  Aug. 26-30, 2013, A. M. University, Pozna\'n, Poland},
  \bibinfo{publisher}{Adam Mickiewicz Univ.\ Press}. pp.
  \bibinfo{pages}{87--101}.
\bibitem[{Kinsman(2015)}]{kinsman15}
\bibinfo{author}{Kinsman, J.H.}, \bibinfo{year}{2015}.
\newblock \bibinfo{title}{A rationale for the initial date of the temple xix
  platform at palenque}.
\newblock \bibinfo{journal}{The Codex, at the University of Pennsylvania Museum
  of Archaeology and Anthropology} \bibinfo{volume}{23},
  \bibinfo{pages}{39--58}.
\bibitem[{Kinsman(2016)}]{kinsman16}
\bibinfo{author}{Kinsman, J.H.}, \bibinfo{year}{2016}.
\newblock \bibinfo{title}{Palenque rulers and mythological time: Evidence of
  sidereal earth year calculations}.
\newblock \bibinfo{note}{Unpublished manuscript in possession of authors}.
\bibitem[{Kohler(2002)}]{kohler02}
\bibinfo{author}{Kohler, U.}, \bibinfo{year}{2002}.
\newblock \bibinfo{title}{Meteors and Comets in Ancient Mexico}.
  \bibinfo{publisher}{Geological Society of America},
  \bibinfo{address}{Boulder, Colorado}.
\newblock pp. \bibinfo{pages}{1--6}.
\newblock \bibinfo{note}{Special Paper 356}.
\bibitem[{Kondrat'eva and Reznikov(1985)}]{kr85}
\bibinfo{author}{Kondrat'eva, E.D.}, \bibinfo{author}{Reznikov, E.A.},
  \bibinfo{year}{1985}.
\newblock \bibinfo{title}{{Comet Tempel-Tuttle and the Leonid meteor swarm}}.
\newblock \bibinfo{journal}{Sol.\ Syst.\ Res.} \bibinfo{volume}{19},
  \bibinfo{pages}{96--101}.
\bibitem[{Landa(1566)}]{landa66}
\bibinfo{author}{Landa, D.d.}, \bibinfo{year}{1566}.
\newblock \bibinfo{title}{Landa's Relacion de las cosas de Yucatan}.
\newblock \bibinfo{publisher}{Peabody Museum of American Archaeology and
  Ethnology, Harvard University}.
\newblock \bibinfo{note}{Translated by Tozzer, 1941}.
\bibitem[{Looper(2003)}]{looper03}
\bibinfo{author}{Looper, M.}, \bibinfo{year}{2003}.
\newblock \bibinfo{title}{Lightning Warrior: Maya Art and Kingship at
  Quirigua}.
\newblock \bibinfo{publisher}{University of Texas Press}.
\bibitem[{Lyytinen et~al.(2001)Lyytinen, Nissinen and van
  Flandern}]{lyytinen01}
\bibinfo{author}{Lyytinen, E.}, \bibinfo{author}{Nissinen, M.},
  \bibinfo{author}{van Flandern, T.}, \bibinfo{year}{2001}.
\newblock \bibinfo{title}{Improved 2001 {L}eonid storm predictions from a
  refined model}.
\newblock \bibinfo{journal}{WGN, J. International Meteor Organization}
  \bibinfo{volume}{29}, \bibinfo{pages}{110--118}.
\bibitem[{Martin(1999)}]{martin99}
\bibinfo{author}{Martin, S.}, \bibinfo{year}{1999}.
\newblock \bibinfo{title}{{The Queen of Middle Classic Tikal}}.
\newblock \bibinfo{journal}{P.A.R.I. Online Publications: Newsletter}
  \bibinfo{volume}{27}, \bibinfo{pages}{1--7}.
\bibitem[{Martin(2003)}]{martin03}
\bibinfo{author}{Martin, S.}, \bibinfo{year}{2003}.
\newblock \bibinfo{title}{In line of the founder: A view of dynastic politics
  at {T}ikal}, in: \bibinfo{editor}{Sabloff, J.A.} (Ed.),
  \bibinfo{booktitle}{Tikal: Dynasties, Foreigners, and Affairs of State,
  Advancing Maya Archaeology}. \bibinfo{publisher}{School of American Research
  Press}. School of American Research Advanced Seminar Series.
  chapter~\bibinfo{chapter}{1}, pp. \bibinfo{pages}{3--45}.
\bibitem[{Martin and Grube(2008)}]{mg00}
\bibinfo{author}{Martin, S.}, \bibinfo{author}{Grube, N.},
  \bibinfo{year}{2008}.
\newblock \bibinfo{title}{Chronicle of the Maya Kings and Queens: Deciphering
  the Dynasties of the Ancient Maya}.
\newblock \bibinfo{edition}{2nd} ed., \bibinfo{publisher}{Thames and Hudson}.
\newblock \bibinfo{note}{(1st ed.\ 2000)}.
\bibitem[{Martin and Skidmore(2012)}]{ms12}
\bibinfo{author}{Martin, S.}, \bibinfo{author}{Skidmore, J.},
  \bibinfo{year}{2012}.
\newblock \bibinfo{title}{Exploring the 584286 correlation between the {M}aya
  and {E}uropean calendars}.
\newblock \bibinfo{journal}{The P.A.R.I. Journal} \bibinfo{volume}{13(2)},
  \bibinfo{pages}{3--16}.
\bibitem[{Maslov(2011)}]{maslov11}
\bibinfo{author}{Maslov, M.}, \bibinfo{year}{2011}.
\newblock \bibinfo{title}{Future {D}raconid outbursts (2011 -- 2100)}.
\newblock \bibinfo{journal}{WGN, J. International Meteor Organization}
  \bibinfo{volume}{39}, \bibinfo{pages}{64--67}.
\bibitem[{Mathews(2016)}]{mathews16}
\bibinfo{author}{Mathews, P.}, \bibinfo{year}{2016}.
\newblock \bibinfo{title}{{The Maya Dates Project}}.
\newblock \bibinfo{note}{Ongoing database of dates from Classic Maya monuments
  and inscriptions. Unpublished.}
\bibitem[{McNaught and Asher(1999)}]{ma99}
\bibinfo{author}{McNaught, R.H.}, \bibinfo{author}{Asher, D.J.},
  \bibinfo{year}{1999}.
\newblock \bibinfo{title}{Leonid dust trails and meteor storms}.
\newblock \bibinfo{journal}{WGN, J. International Meteor Organization}
  \bibinfo{volume}{27}, \bibinfo{pages}{85--102}.
\bibitem[{Meeus(2000)}]{meeus00}
\bibinfo{author}{Meeus, J.}, \bibinfo{year}{2000}.
\newblock \bibinfo{title}{Astronomical Algorithms}.
\newblock \bibinfo{edition}{Second {E}nglish} ed.,
  \bibinfo{publisher}{Willmann-Bell, Inc.}
\bibitem[{Murray and Dermott(1999)}]{md99}
\bibinfo{author}{Murray, C.D.}, \bibinfo{author}{Dermott, S.F.},
  \bibinfo{year}{1999}.
\newblock \bibinfo{title}{Solar System Dynamics}.
\newblock \bibinfo{publisher}{Cambridge University Press}.
\bibitem[{Pankenier et~al.(2008)Pankenier, Xu and Jiang}]{pxj08}
\bibinfo{author}{Pankenier, D.W.}, \bibinfo{author}{Xu, Z.},
  \bibinfo{author}{Jiang, Y.}, \bibinfo{year}{2008}.
\newblock \bibinfo{title}{Archaeoastronomy in East Asia: Historical
  Observational Records of Comets and Meteor Showers from China, Japan, and
  Korea}.
\newblock \bibinfo{publisher}{Cambria Press}.
\bibitem[{{Peale}(1976)}]{peale76}
\bibinfo{author}{{Peale}, S.J.}, \bibinfo{year}{1976}.
\newblock \bibinfo{title}{{Orbital resonances in the solar system}}.
\newblock \bibinfo{journal}{Ann.\ Rev.\ Astron.\ Astrophys.}
  \bibinfo{volume}{14}, \bibinfo{pages}{215--246}.
\bibitem[{Plavec(1956)}]{plavec56}
\bibinfo{author}{Plavec, M.}, \bibinfo{year}{1956}.
\newblock \bibinfo{title}{On the evolution of the meteor streams}.
\newblock \bibinfo{journal}{Vistas Astron.} \bibinfo{volume}{2},
  \bibinfo{pages}{994--998}.
\bibitem[{Plavec(1957)}]{plavec57}
\bibinfo{author}{Plavec, M.}, \bibinfo{year}{1957}.
\newblock \bibinfo{title}{On the origin and early stages of the meteor
  streams}.
\newblock \bibinfo{journal}{Publ.\ Astron.\ Inst.\ Czechosl.\ Acad.\ Sci.}
  \bibinfo{volume}{30}, \bibinfo{pages}{1--94}.
\bibitem[{Powell(1997)}]{powell97}
\bibinfo{author}{Powell, C.}, \bibinfo{year}{1997}.
\newblock \bibinfo{title}{A New View on Maya Astronomy}.
\newblock Master's thesis. University of Texas at Austin.
\bibitem[{{Rendtel}(2007)}]{rendtel07}
\bibinfo{author}{{Rendtel}, J.}, \bibinfo{year}{2007}.
\newblock \bibinfo{title}{{Three days of enhanced Orionid activity in 2006 --
  Meteoroids from a resonance region?}}
\newblock \bibinfo{journal}{WGN, J. International Meteor Organization}
  \bibinfo{volume}{35}, \bibinfo{pages}{41--45}.
\bibitem[{{Rendtel}(2008)}]{rendtel08}
\bibinfo{author}{{Rendtel}, J.}, \bibinfo{year}{2008}.
\newblock \bibinfo{title}{The {O}rionid meteor shower observed over 70 years}.
\newblock \bibinfo{journal}{Earth Moon Plan.} \bibinfo{volume}{102},
  \bibinfo{pages}{103--110}.
\bibitem[{{Reznikov}(1983)}]{reznikov83}
\bibinfo{author}{{Reznikov}, E.A.}, \bibinfo{year}{1983}.
\newblock \bibinfo{title}{{Origin of the Bootid meteoroid shower}}.
\newblock \bibinfo{journal}{Trudy Kazanskaia Gorodkoj Astron.\ Obs.}
  \bibinfo{volume}{47}, \bibinfo{pages}{131--136}.
\newblock \bibinfo{note}{In Russian}.
\bibitem[{{Sato} and {Watanabe}(2007)}]{sw07}
\bibinfo{author}{{Sato}, M.}, \bibinfo{author}{{Watanabe}, J.},
  \bibinfo{year}{2007}.
\newblock \bibinfo{title}{{Origin of the 2006 Orionid outburst}}.
\newblock \bibinfo{journal}{Publ.\ Astron.\ Soc.\ Japan} \bibinfo{volume}{59},
  \bibinfo{pages}{L21--L24}.
\bibitem[{Sato and Watanabe(2010)}]{sw10}
\bibinfo{author}{Sato, M.}, \bibinfo{author}{Watanabe, J.},
  \bibinfo{year}{2010}.
\newblock \bibinfo{title}{Forecast for {P}hoenicids in 2008, 2014, and 2019}.
\newblock \bibinfo{journal}{Publ.\ Astron.\ Soc.\ Japan} \bibinfo{volume}{62},
  \bibinfo{pages}{509--513}.
\bibitem[{Sato and Watanabe(2014)}]{sw14}
\bibinfo{author}{Sato, M.}, \bibinfo{author}{Watanabe, J.},
  \bibinfo{year}{2014}.
\newblock \bibinfo{title}{Forecast of enhanced activity of eta-{A}quariids in
  2013}, in: \bibinfo{editor}{Jopek, T.J.}, \bibinfo{editor}{Rietmeijer,
  F.J.M.}, \bibinfo{editor}{Watanabe, J.}, \bibinfo{editor}{Williams, I.P.}
  (Eds.), \bibinfo{booktitle}{Proceedings of the Meteoroids 2013 Conference,
  Aug. 26-30, 2013, A. M. University, Pozna\'n, Poland},
  \bibinfo{publisher}{Adam Mickiewicz Univ.\ Press}. pp.
  \bibinfo{pages}{213--216}.
\bibitem[{Saturno et~al.(2012)Saturno, Stuart, Aveni and Rossi}]{ssar12}
\bibinfo{author}{Saturno, W.A.}, \bibinfo{author}{Stuart, D.},
  \bibinfo{author}{Aveni, A.F.}, \bibinfo{author}{Rossi, F.},
  \bibinfo{year}{2012}.
\newblock \bibinfo{title}{Ancient {M}aya astronomical tables from {Xultun,
  Guatemala}}.
\newblock \bibinfo{journal}{Science} \bibinfo{volume}{336},
  \bibinfo{pages}{714--717}.
\bibitem[{Schele et~al.(1992)Schele, Grube and Fahsen}]{sgf92}
\bibinfo{author}{Schele, L.}, \bibinfo{author}{Grube, N.},
  \bibinfo{author}{Fahsen, F.}, \bibinfo{year}{1992}.
\newblock \bibinfo{title}{The Lunar Series in Classic Maya Inscriptions}.
\newblock \bibinfo{type}{Technical Report} \bibinfo{number}{29}. The CHAAAC of
  the Art Department of the University of Texas at Austin.
\newblock \bibinfo{note}{Texas Notes on Precolumbian Art, Writing, and
  Culture}.
\bibitem[{Seargent(2009)}]{seargent09}
\bibinfo{author}{Seargent, D.}, \bibinfo{year}{2009}.
\newblock \bibinfo{title}{The Greatest Comets in History: Broom Stars and
  Celestial Scimitars}.
\newblock \bibinfo{publisher}{Springer}.
\bibitem[{Sekhar and Asher(2013)}]{sa13}
\bibinfo{author}{Sekhar, A.}, \bibinfo{author}{Asher, D.J.},
  \bibinfo{year}{2013}.
\newblock \bibinfo{title}{Saturnian mean motion resonances in meteoroid
  streams}.
\newblock \bibinfo{journal}{Mon.\ Not.\ Roy.\ Astron.\ Soc.}
  \bibinfo{volume}{433}, \bibinfo{pages}{L84--L88}.
\bibitem[{Sekhar and Asher(2014)}]{sa14}
\bibinfo{author}{Sekhar, A.}, \bibinfo{author}{Asher, D.J.},
  \bibinfo{year}{2014}.
\newblock \bibinfo{title}{{Resonant behavior of Comet Halley and the Orionid
  stream}}.
\newblock \bibinfo{journal}{Meteorit.\ Planet.\ Sci.} \bibinfo{volume}{49},
  \bibinfo{pages}{52--62}.
\bibitem[{{Sekhar} et~al.(2016){Sekhar}, {Asher} and {Vaubaillon}}]{sekhar16}
\bibinfo{author}{{Sekhar}, A.}, \bibinfo{author}{{Asher}, D.J.},
  \bibinfo{author}{{Vaubaillon}, J.}, \bibinfo{year}{2016}.
\newblock \bibinfo{title}{{Three-body resonance in meteoroid streams}}.
\newblock \bibinfo{journal}{Mon.\ Not.\ Roy.\ Astron.\ Soc.}
  \bibinfo{volume}{460}, \bibinfo{pages}{1417--1427}.
\bibitem[{Stephenson et~al.(1985)Stephenson, Yau and Hunger}]{stephenson85}
\bibinfo{author}{Stephenson, F.R.}, \bibinfo{author}{Yau, K.K.C.},
  \bibinfo{author}{Hunger, H.}, \bibinfo{year}{1985}.
\newblock \bibinfo{title}{Records of halley's comet on babylonian tablets}.
\newblock \bibinfo{journal}{Nature} \bibinfo{volume}{314},
  \bibinfo{pages}{587--592}.
\bibitem[{Stuart(2000)}]{stuart00}
\bibinfo{author}{Stuart, D.}, \bibinfo{year}{2000}.
\newblock \bibinfo{title}{The arrival of strangers: {T}eotihuacan and {T}ollan
  in classic {M}aya history}, in: \bibinfo{editor}{Carrasco, D.},
  \bibinfo{editor}{Jones, L.}, \bibinfo{editor}{Sessions, S.} (Eds.),
  \bibinfo{booktitle}{Mesoamerica's Classic Heritage: from Teotihuacan to the
  Aztecs}. \bibinfo{edition}{2002} ed.. \bibinfo{publisher}{University Press of
  Colorado}. chapter~\bibinfo{chapter}{15}, pp. \bibinfo{pages}{465--513}.
\bibitem[{Stuart(2005)}]{stuart05}
\bibinfo{author}{Stuart, D.}, \bibinfo{year}{2005}.
\newblock \bibinfo{title}{The Inscriptions from Temple XIX at Palenque: a
  Commentary}.
\newblock \bibinfo{publisher}{The Pre-Columbian Art Research Institute}.
\bibitem[{Stuart(2007)}]{stuart07}
\bibinfo{author}{Stuart, D.}, \bibinfo{year}{2007}.
\newblock \bibinfo{title}{Hit the road}.
\newblock \bibinfo{howpublished}{Electronic document online at
  decipherment.wordpress.com}.
\newblock \bibinfo{note}{Maya Decipherment: Ideas on Ancient Maya Writing and
  Iconography}.
\bibitem[{Taube(2000)}]{taube00}
\bibinfo{author}{Taube, K.}, \bibinfo{year}{2000}.
\newblock \bibinfo{title}{The turquoise hearth: Fire, self-sacrifice, and the
  central {M}exican cult of war}, in: \bibinfo{editor}{Carrasco, D.},
  \bibinfo{editor}{Jones, L.}, \bibinfo{editor}{Sessions, S.} (Eds.),
  \bibinfo{booktitle}{Mesoamerica's Classic Heritage: from Teotihuacan to the
  Aztecs}. \bibinfo{edition}{2002} ed.. \bibinfo{publisher}{University Press of
  Colorado}. chapter~\bibinfo{chapter}{10}, pp. \bibinfo{pages}{269--340}.
\bibitem[{Telleriano-Remensis(1901)}]{famsi16a}
\bibinfo{author}{Telleriano-Remensis}, \bibinfo{year}{1901}.
\newblock \bibinfo{title}{Codex telleriano-remensis}.
\newblock \bibinfo{howpublished}{Foundation for the Advancement of Mesoamerican
  Studies, Inc.}
\newblock \bibinfo{note}{Page 39V}.
\bibitem[{Trenary(1987-1988)}]{trenary87}
\bibinfo{author}{Trenary, C.}, \bibinfo{year}{1987-1988}.
\newblock \bibinfo{title}{Universal meteor metaphors and their occurrence in
  {M}esoamerican astronomy}.
\newblock \bibinfo{journal}{Archaeoastronomy} \bibinfo{volume}{10},
  \bibinfo{pages}{99--116}.
\bibitem[{Van~Laningham()}]{pau16}
\bibinfo{author}{Van~Laningham, I.}, .
\newblock \bibinfo{title}{pauahtun.org}.
\bibitem[{Vaticanus-3773()}]{famsi16c}
\bibinfo{author}{Vaticanus-3773}, .
\newblock \bibinfo{title}{Codex vaticanus 3773}.
\newblock \bibinfo{howpublished}{Foundation for the Advancement of Mesoamerican
  Studies, Inc.}
\newblock \bibinfo{note}{Facsimile, Electronic document,
  http://www/famsi/org/research/codices}.
\bibitem[{Whipple(1951)}]{whipple51}
\bibinfo{author}{Whipple, F.L.}, \bibinfo{year}{1951}.
\newblock \bibinfo{title}{A comet model. {II. P}hysical relations for comets
  and meteors}.
\newblock \bibinfo{journal}{Astrophys.\ J.} \bibinfo{volume}{113},
  \bibinfo{pages}{464--474}.
\bibitem[{Yeomans and Kiang(1981)}]{yk81}
\bibinfo{author}{Yeomans, D.K.}, \bibinfo{author}{Kiang, T.},
  \bibinfo{year}{1981}.
\newblock \bibinfo{title}{The long-term motion of comet {H}alley}.
\newblock \bibinfo{journal}{Mon.\ Not.\ Roy.\ Astron.\ Soc.}
  \bibinfo{volume}{197}, \bibinfo{pages}{633--646}.
\bibitem[{Zhuang(1977)}]{zhuang77}
\bibinfo{author}{Zhuang, T.S.}, \bibinfo{year}{1977}.
\newblock \bibinfo{title}{Ancient {C}hinese reports of meteor showers}.
\newblock \bibinfo{journal}{Chinese Astron.} \bibinfo{volume}{1},
  \bibinfo{pages}{197--220}.

\end{thebibliography}

\appendix

\section{ETA Data Set}
\label{app-data}

\begin{tabbing}
967 BC Apr  7  \=  5.8.17.15.17 \0\0 \= PAL \0\0 \= acc (U Kokan Chan) \\
249    Apr \07 \>  8.10.10.10.16  \> CPN \> unk \\
328    Apr 11  \>  8.14.10.13.15  \> WAX \> unk \\
480    Apr 18  \>  9.2.5.0.0      \> QRG \> pe \\
484    Apr 13  \>  9.2.9.0.16     \> CRC \> acc (Yajaw Te' K'inich I) \\
511    Apr 20  \>  9.3.16.8.4     \> TIK \> acc (Lady of Tikal) \\
531    Apr 14  \>  9.4.16.13.3    \> CRC \> acc (K'an I) \\
553    Apr 17  \>  9.5.19.1.2     \> CRC \> acc (Yajaw Te' K'inich II) \\
556    Apr 10  \>  9.6.2.1.11     \> CRC \> war (axing) \\
562    Apr 30  \>  9.6.8.4.2      \> CRC \> war (Star War) \\
566    Apr 23  \>  9.6.12.4.16    \> CRC \> birth (Lady B'atz??? Ek') \\
572    Apr \07 \>  9.6.18.5.12    \> PAL \> acc  (Kan B'ahlam I) \\
588    Apr 19  \>  9.7.14.10.8    \> CRC \> birth (K'an II) \\
599    Apr 13  \>  9.8.5.12.19    \> TNA \> unk \\
606    Apr 14  \>  9.8.12.14.17   \> TNA \> birth (Hix Chapat) \\
611    Apr \05 \>  9.8.17.15.14   \> PAL \> war (axing) \\
614    Apr 12  \>  9.9.0.16.17    \> CRC \> tomb \\
618    Apr 14  \>  9.9.5.0.0      \> ALS \> pe \\
621    Apr \09 \>  9.9.8.0.11     \> PNG \> unk \\
634    Apr \09 \>  9.10.1.3.19    \> DPL \> depart (B'alaj Chan K'awiil) \\
635    Apr \09 \>  9.10.2.4.4     \> CRN \> foundation? \\
636    Apr 29  \>  9.10.3.5.10    \> CLK \> acc (Yuknoom Ch'en II) \\
639    Apr 13  \>  9.10.6.5.9     \> PNG \> acc (Ruler 2) \\
640    Apr \07 \>  9.10.7.5.9     \> COB \> acc (Ruler A) \\
644    Apr \09 \>  9.10.11.6.12   \> NAR \> strike (stone road) \\
662    Apr \06 \>  9.11.9.11.3    \> MRL \> acc (2nd, Muwan Jol? Pakal) \\
663    Apr 23  \>  9.11.10.12.5   \> CRN \> dedication (house) \\
675    Apr 26  \>  9.12.2.15.11   \> CRN \> depart (CLK king) \\
686    Apr \04 \>  9.12.13.17.7   \> CLK \> acc (Yuknoom Yich'aak K'ahk') \\
687    Apr 12  \>  9.12.15.0.0    \> PNG \> pe \\
690    Apr \09 \>  9.12.18.0.13   \> CRN \> fire \\
691    Apr 12  \>  9.12.19.1.1    \> TZE \> fire (tomb) \\
694    Apr 23  \>  9.13.2.2.8     \> CRN \> unk (Chak Ak'ach Yuk) \\
699    Apr 17  \>  9.13.7.3.8     \> NAR \> dedication (ceremonies) \\
711    Apr \09 \>  9.13.19.6.3    \> NAR \> unk \\
716    Apr \04 \>  9.14.3.8.4     \> NAR \> war (attack) \\
721    Apr 27  \>  9.14.9.9.14    \> CRN \> arrive (`Ix Ti' Kan Ajaw) \\
723    Apr 24  \>  9.14.11.10.1   \> YAX \> fire \\
726    Apr \05 \>  9.14.14.9.18   \> PNG \> unk \\
726    Apr 21  \>  9.14.14.10.14  \> NAR \> war? (star war?) \\
738    Apr \05 \>  9.15.6.13.1    \> YAX \> fire \\
738    Apr 24  \>  9.15.6.14.0    \> QRG \> fire \\
738    Apr 30  \>  9.15.6.14.6    \> QRG \> war (decapitation) \\
750    Apr \08 \>  9.15.18.16.7   \> PNG \> birth (Ruler 7) \\
752    Apr 30  \>  9.16.1.0.0     \> YAX \> acc, pe (Bird Jaguar IV) \\
756    Apr 10  \>  9.16.5.0.0     \> multiple sites \0\0 pe \\
770    Apr \09 \>  9.16.19.3.12   \> EKB \> arrive (Ukit Kan Lek) \\
778    Apr 13  \>  9.17.7.5.19    \> AGT \> war (downing of ``flint-shield'' [army]) \\
778    Apr  17 \>  9.17.7.6.3     \> ITZ \> smoke \\
780    Apr 13  \>  9.17.9.6.14    \> IXK \> dedication \\
781    Apr 18  \>  9.17.10.7.0    \> HIG \> acc (unk) \\
783    Apr 16  \>  9.17.12 .7.8   \> QRG \> unk \\
790    Apr 12  \>  9.17.19.9.1    \> NAR \> strike (stone road) \\
790    Apr 26  \>  9.17.19.9.15   \> QRG \> unk \\
796    Apr \02 \>  9.18.5.10.3    \> TNA \> death (Aj Tolol Te')  \\
802    Apr \08 \>  9.18.11.12.0   \> AML \> acc (Lachan K'awiil Ajaw Bot) \\
808    Apr \06 \>  9.18.17.13.10  \> YAX \> fire \\
808    Apr 10  \>  9.18.17.13.14  \> YAX \> throw \\
820    Apr 13  \>  9.19.9.17.0    \> CRC \> tomb? \\
849    Apr 15  \>  10.0.19.6.14   \> CRC \> ???form the earth???? \\
863    Apr 12  \>  10.1.13.10.4   \> Randel Stela \0\0 819 day count \\
\end{tabbing}

\end{document}